%% file: DESY-07-100.tex
\documentclass[zpreprint,zbstnp]{zeus_paper}
\usepackage[english]{babel}

\input zeus_def.tex

\def\citeCTD{{\cite{%
nim:a279:290,*npps:b32:181,*nim:a338:254%
}}\xspace}
\def\citeCAL{{\cite{%
nim:a309:77,*nim:a309:101,*nim:a321:356,*nim:a336:23%
}}\xspace}

\begin{document}

\prepnum{{DESY--07--100}}

\title{
Forward-jet production in deep inelastic $ep$ scattering at HERA}

\author{ZEUS Collaboration}
\date{July 2007}

\abstract{
Forward jet cross sections have been measured in neutral
current deep inelastic scattering at low Bjorken-$x$
with the ZEUS detector at HERA using an
integrated luminosity of ${81.8~\rm pb}^{-1}$.
Measurements are presented for inclusive forward jets as
well as for forward jets accompanied by a dijet system.
The explored phase space, with jet pseudorapidity up to 4.3
is expected to be particularly sensitive to the dynamics of
QCD parton evolution at low $x$. The
measurements are compared to fixed-order QCD
calculations and to leading-order parton-shower Monte
Carlo models.
}

\makezeustitle

\def\3{\ss}

\pagenumbering{Roman}

\begin{center}
{                      \Large  The ZEUS Collaboration              }
\end{center}
  S.~Chekanov$^{   1}$,                                                                            
  M.~Derrick,                                                                                      
  S.~Magill,                                                                                       
  B.~Musgrave,                                                                                     
  D.~Nicholass$^{   2}$,                                                                           
  \mbox{J.~Repond},                                                                                
  R.~Yoshida\\                                                                                     
 {\it Argonne National Laboratory, Argonne, Illinois 60439-4815}, USA~$^{n}$                       
\par \filbreak                                                                                     
  M.C.K.~Mattingly \\                                                                              
 {\it Andrews University, Berrien Springs, Michigan 49104-0380}, USA                               
\par \filbreak                                                                                     
  M.~Jechow, N.~Pavel~$^{\dagger}$, A.G.~Yag\"ues Molina \\                                        
  {\it Institut f\"ur Physik der Humboldt-Universit\"at zu Berlin,                                 
           Berlin, Germany}
% ...
\newpage                                                                                           
$^{\    1}$ supported by DESY, Germany \\                                                          
$^{\    2}$ also affiliated with University College London, UK \\                                  
% ...
$^{\dagger}$ deceased \\                                                                           
%                                                                                                  
% \par         % if index listing & table fit to 1 page, put gap here                              
%\newpage   % alternatively: go to newpage, if page is too small                                    
                                                           %                                       
% \institute_references_start    % do not touch or move this line !                                
                                                           %                                       
\begin{tabular}[h]{rp{14cm}}                                                                       
$^{a}$ &  supported by the Natural Sciences and Engineering Research Council of Canada (NSERC) \\  
$^{b}$ &  supported by the German Federal Ministry for Education and Research (BMBF), under        
          contract numbers HZ1GUA 2, HZ1GUB 0, HZ1PDA 5, HZ1VFA 5\\                                
$^{c}$ &  supported in part by the MINERVA Gesellschaft f\"ur Forschung GmbH, the Israel Science   
          Foundation (grant no. 293/02-11.2) and the U.S.-Israel Binational Science Foundation \\  
$^{d}$ &  supported by the German-Israeli Foundation and the Israel Science Foundation\\           
$^{e}$ &  supported by the Italian National Institute for Nuclear Physics (INFN) \\                
$^{f}$ &  supported by the Japanese Ministry of Education, Culture, Sports, Science and Technology 
          (MEXT) and its grants for Scientific Research\\                                          
$^{g}$ &  supported by the Korean Ministry of Education and Korea Science and Engineering          
          Foundation\\                                                                             
$^{h}$ &  supported by the Netherlands Foundation for Research on Matter (FOM)\\                   
$^{i}$ &  supported by the Polish State Committee for Scientific Research, grant no.               
          620/E-77/SPB/DESY/P-03/DZ 117/2003-2005 and grant no. 1P03B07427/2004-2006\\             
$^{j}$ &  partially supported by the German Federal Ministry for Education and Research (BMBF)\\   
$^{k}$ &  supported by RF Presidential grant N 8122.2006.2 for the leading                         
          scientific schools and by the Russian Ministry of Education and Science through its grant
          Research on High Energy Physics\\                                                        
$^{l}$ &  supported by the Spanish Ministry of Education and Science through funds provided by     
          CICYT\\                                                                                  
$^{m}$ &  supported by the Particle Physics and Astronomy Research Council, UK\\                   
$^{n}$ &  supported by the US Department of Energy\\                                               
$^{o}$ &  supported by the US National Science Foundation. Any opinion,                            
findings and conclusions or recommendations expressed in this material                             
are those of the authors and do not necessarily reflect the views of the                           
National Science Foundation.\\                                                                     
$^{p}$ &  supported by the Polish Ministry of Science and Higher Education                         
as a scientific project (2006-2008)\\                                                              
$^{q}$ &  supported by FNRS and its associated funds (IISN and FRIA) and by an Inter-University    
          Attraction Poles Programme subsidised by the Belgian Federal Science Policy Office\\     
$^{r}$ &  supported by the Malaysian Ministry of Science, Technology and                           
Innovation/Akademi Sains Malaysia grant SAGA 66-02-03-0048\\                                       
\end{tabular}                                                                                      

\newpage

% ----------------------------------------------------------------------------
%       Introduction
% ----------------------------------------------------------------------------
\section{Introduction}
\label{sec-int}

Deep inelastic lepton scattering (DIS) off protons provides a rich field
for exploring the parton dynamics in QCD. HERA has extended the
phase-space region in the Bjorken scaling variable, $x_{\rm Bj}$, down to
a few $10^{-5}$. At such low $x_{\rm Bj}$, several steps in the QCD
cascade initiated by a parton from the proton can occur before the final
interaction with the virtual photon takes place. The result of this
cascade may be observed in the final state and provides an opportunity
to study the QCD parton evolution in detail.  

Within perturbative QCD (pQCD), fixed-order calculations for the parton
evolution are so far available only at next-to-leading order (NLO). 
A number of different approximations to the QCD evolution have been
developed, based on summing of particular subsets of diagrams in accordance
with their importance in the phase space considered.

The conventional DGLAP~\cite{sovjnp:15:438,jetp:46:641,np:b126:298} approach
sums up the leading logarithms in the virtuality of the exchanged boson,
$Q^2$, and is expected to be valid at not too small $x_{\rm Bj}$ and $Q^2$.  At
small $x_{\rm Bj}$, a better approximation is expected to be provided by the
BFKL formalism~\cite{sovjnp:28:822} which resums the leading logarithm terms
in $1/x$, where $x$ is the fractional longitudinal momentum of a parton. The
CCFM~\cite{np:b296:49,np:b336:18} approach interpolates between the two types
of evolution, DGLAP and BFKL.

The DGLAP evolution equations have been successfully
tested at HERA in inclusive measurements at low $x_{\rm Bj}$ and no indication
of BFKL dynamics was observed. 
The dynamics at low $x_{\rm Bj}$ can be further probed by
measurements of the partonic final state that highlight the
differences between predictions of the BFKL and DGLAP formalisms. BFKL
evolution results in a larger fraction of small $x_{\rm Bj}$ events with
forward jets\footnote{The ZEUS
  coordinate system is a right-handed Cartesian system, with the $Z$
  axis pointing in the proton beam direction, referred to as the
  "forward direction", and the $X$ axis pointing towards the centre of
  HERA.  The coordinate origin is at the nominal interaction point.}
than the DGLAP evolution. A forward jet is characterised 
by a high fractional longitudinal momentum, $x^{\rm jet}=p_{Z}^{\rm jet}/p$, 
where $p$ is the proton momentum and $p_{\rm Z}^{\rm jet}$ is the 
longitudinal jet momentum~\cite{npps:18c:125}. 

A comparison of data on forward jets with the DGLAP leading-order parton-shower Monte
Carlo programmes performed previously in DIS at
HERA~\cite{epj:c6:239,pl:b474:223,np:b538:3} has revealed a clear deficit of forward
jets in the Monte Carlo. However, the addition of a parton cascade evolved
according to DGLAP on the photon side has significantly improved
the description ~\cite{cpc:86:147}. The
simulation based on the Color Dipole Model
(CDM)~\cite{pl:b175:453,np:b306:746,zfp:c43:625}, which includes parton emissions not
ordered in transverse momentum, also succeeded in describing the data. Fixed-order NLO QCD
calculations~\cite{np:b485:291,nlojet} were also compared to the forward jets measurements
in more recent
publications and failed to describe the data. These studies were performed up to
pseudorapidities 3~\cite{desy-05-117} and 2.7~\cite{desy-05-135}.

In this paper, measurements of inclusive forward-jet cross sections for
pseudorapidities of up to 4.3 are presented, based on a data sample which
corresponds to a twofold increase in luminosity with respect to the previous ZEUS 
analysis~\cite{desy-05-117}. Furthermore, a comparison of the measured cross
sections with the {\sc Cascade} Monte Carlo~\cite{epj:c19:351}, based on the CCFM
evolution, is presented. In addition, measurements of ``forward jet + dijet'' cross
sections, as investigated by the H1 Collaboration~\cite{desy-05-135}, are
reported. These measurements explore parton evolution in a more exclusive way and
are more sensitive to its details.

% ----------------------------------------------------------------------------
%      Section 2  Theoretical framework and phase-space definitions 
% ----------------------------------------------------------------------------
\section{Theoretical framework and phase-space definitions}
\label{sec-theory}

The DGLAP evolution equations, based on collinear factorisation,
assume that the dominant contribution to parton evolution comes
from subsequent parton emissions that are strongly ordered in
transverse momenta, $k_T$, where the parton with the largest $k_T$
interacts with the photon. In this formalism, only the leading terms
in  $\ln Q^2$ in the QCD perturbative expansion are summed up. Since this
approximation does not resum leading $\ln 1/x$ terms, it may not be adequate at low
$x_{\rm Bj}$.

Contrary to the DGLAP approach, the BFKL evolution equation resums the leading 
$\ln 1/x$ and the evolution proceeds over $x$ at fixed $Q^2$. The BFKL
approach imposes no ordering in $k_T$ but strong ordering in $x$, with the low-$x$
parton interacting with the photon.  At small $x_{\rm Bj}$, the difference
between these approaches to the QCD parton evolution is expected to be most
prominent for hard partons created at the beginning of the cascade, i.e. at
pseudorapidities close to the proton (forward region).

The CCFM approach is based on the idea of coherent gluon radiation, which leads to
angular ordering of gluon emissions in the gluon ladder. It interpolates between
the above two types of evolution, so it should be applicable over a larger
phase-space region.

A phenomenological approach to parton evolution is provided by CDM~\cite{pl:b175:453,np:b306:746,zfp:c43:625}.
In this model, gluons are emitted by color dipoles successively spanned between partons in the
cascade. Due to independent radiation of the dipoles, the emitted gluons are not
ordered in $k_T$ and thus CDM mimics the BFKL-type evolution.  

%%%%%%%%%%%%%%%%%%%%%%%%%%%%%%%%%%%%%%%%%%%%%%%%%%%%%%%%%%%%%%%%%%%%%%%%%%%%%

To investigate the relevance of different approaches, events
with a jet in the forward region were analysed in the low-$x_{\rm Bj}$
region. Events were required to have at least one jet satisfying the
following criteria:

\begin{itemize}
\item $(p_T^{\rm jet})^2 \sim Q^2$;
\item $x^{\rm jet} \gg x_{\rm Bj}$,
\end{itemize}

where $p_T^{\rm jet}$ is the transverse momentum of the jet. The first
condition suppresses strong ordering in the transverse momenta and
decreases the probability of having a DGLAP-type evolution. The second
condition enhances the phase space for the BFKL evolution.

A further event sample called ``forward jet+dijet'', which contains at least
two hard jets in addition to the forward jet (fjet), was selected. The jets were
ordered in pseudorapidity such that
${\eta}^{\rm jet_1} < {\eta}^{\rm jet_2} < {\eta}^{\rm fjet}$. For
this sample, the pseudorapidity separation of dijets,
$\Delta\eta_1={\eta}^{\rm jet_{2}}-{\eta}^{\rm jet_{1}}$, and the
pseudorapidity difference between the forward and the second jet of the dijet, $\Delta\eta_2={\eta}^{\rm fjet}-{\eta}^{\rm jet_{2}}$,
were studied.

\begin{figure}[tbhp]
\begin{center}
\includegraphics[height=80mm]{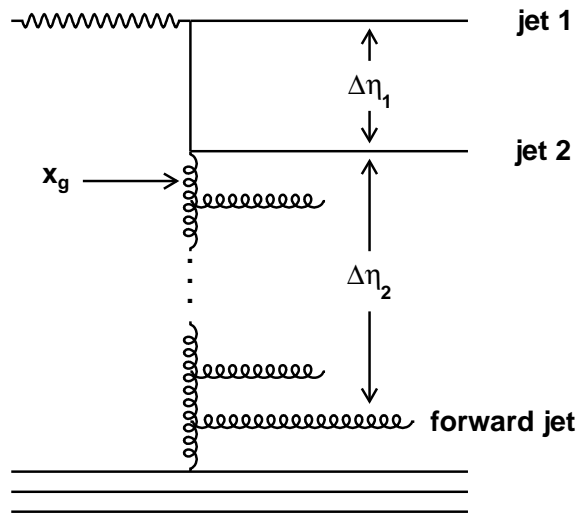}
\end{center}
\caption{A schematic diagram of an interaction in which a forward jet and
two additional hard jets can be produced.
}
\label{fig-dijet}
\end{figure}

The cross section as a function of $\Delta\eta_2$ was investigated for two
intervals of $\Delta\eta_1$, $\Delta\eta_1<1$ and $\Delta\eta_1>1$. With such
a choice of $\Delta\eta_1$, different dynamics of the partons in the cascade
are expected to be highlighted. For $\Delta\eta_1<1$, small invariant masses of
the dijet system are favoured and, therefore, partons with small values of $x_g$ are
produced, where $x_g$ is the longitudinal momentum fraction carried by the gluon
coupled to the hard dijet system (Fig.~\ref{fig-dijet}). Consequently, a large
space is left for BFKL-type evolution in $x$ from the forward jet to the dijet
system.  When $\Delta\eta_1$ is large, BFKL-like evolution can occur between the
partons producing the dijet system.  

% ----------------------------------------------------------------------------
%      Section 3   Experimental set-up
% ----------------------------------------------------------------------------
\section{Experimental set-up}
\label{sec-expset}

The analysis was performed with the data taken with the ZEUS detector
from 1998 to 2000, when HERA
collided electrons or positrons\footnote{Hereafter, both $e^+$ and $e^-$
are referred to as electrons, unless explicitly stated otherwise.}
with energy of $E_e = 27.5\gev$ with protons
of energy $E_p = 920\gev$, yielding a centre-of-mass energy of $318\gev$.
The results are based on the sum of the $e^-p$
and $e^+p$ samples, corresponding to integrated luminosities of
$16.4\pm0.3\,$pb$^{-1}$ and $65.3\pm1.5\,$pb$^{-1}$, respectively.

\Zdetdesc

%\Zctddesc\ZcoosysfnBeta
%\Zcaldesc

%A detailed description of the ZEUS detector can be found elsewhere~\cite{zeus:1993:bluebook}.
%A brief outline of the components that are most relevant for this analysis
%is given below.

\Zctddesc\ZcoosysfnBeta

%Charged particle tracks are reconstructed in the central tracking
%detector (CTD)~\cite{nim:a279:290,*npps:b32:181,*nim:a338:254}, which operates
%in a magnetic field of $1.43\Tesla$ provided by a thin superconducting solenoid.
%The CTD consists of 72~cylindrical drift-chamber layers, organised in nine
%superlayers covering the polar-angle
%region \mbox{$15^\circ<\theta<164^\circ$}. The transverse-momentum resolution
%for full-length tracks can be parameterised as
%$\sigma(p_T)/p_T=0.0058p_T\oplus0.0065\oplus0.0014/p_T$, 
%with $p_T$ in $\Gev$. The tracking system was 
%used to measure the interaction vertex
%with a typical resolution along (transverse to) the beam direction of
%0.4~(0.1)~cm and to cross-check the energy scale of the calorimeter.

 \Zcaldesc

%The high-resolution uranium--scintillator calorimeter
%(CAL)~\cite{nim:a309:77,*nim:a309:101,*nim:a321:356,*nim:a336:23} covers
%$99.7$\% of the total solid angle and consists
%of three parts: the forward (FCAL, $2.6^{\circ}<\theta<36.7^{\circ}$), the
%barrel (BCAL, $36.7^{\circ}<\theta<129.1^{\circ}$) and the rear
%(RCAL, $129.1^{\circ}<\theta<176.2^{\circ}$) calorimeters. Each part is
%subdivided transversely into towers and longitudinally into one electromagnetic
%section (EMC) and either one (in RCAL) or two (in BCAL and FCAL) hadronic
%sections (HAC). The smallest subdivision of the calorimeter is called a cell.
%Under test-beam conditions, the CAL single-particle relative energy
%resolutions were $\sigma(E)/E=0.18/\sqrt E$ for electrons and
%$\sigma(E)/E=0.35/\sqrt E$ for hadrons, with $E$ in GeV.

The luminosity was measured using the bremsstrahlung
process $ep\rightarrow e\gamma p$ with the luminosity
monitor~\cite{desy-92-066,*zfp:c63:391,*acpp:b32:2025},
a lead-scintillator calorimeter placed in the HERA tunnel at $Z=-107$~m.

%The luminosity was measured from the rate of the bremsstrahlung
%process $ep\rightarrow e\gamma p$. The resulting small-angle
%energetic photons were measured by the luminosity
%monitor~\cite{desy-92-066,*zfp:c63:391,*acpp:b32:2025}, a
%lead-scintillator calorimeter placed in the HERA tunnel at $Z=-107$~m.

For the 1998-2000 running period, the forward plug calorimeter 
(FPC)~\cite{fpcnim:a450:235}
was installed in the $20 \times 20$ cm${}^2$ beam hole of the FCAL,
with a small hole of radius $3.15$ cm in the center to accommodate the
beam pipe. The FPC increased the forward calorimetric coverage
by about 1 unit of pseudorapidity to $\eta\,\leq\,5$. The FPC consisted of
a lead--scintillator sandwich calorimeter divided longitudinally into
electromagnetic and hadronic sections that were read out separately by
wavelength-shifting  fibers and photomultipliers. The energy resolution,
as measured under test-beam conditions, was
$\sigma(E)/E = 0.41/\sqrt{E} \oplus 0.062$ and
$\sigma(E)/E = 0.65/\sqrt{E} \oplus 0.06$ for electrons and pions,
respectively, with $E$ in GeV.

% ----------------------------------------------------------------------------
%       Section 4 Event selection and jet definition
% ----------------------------------------------------------------------------
\section{Event selection and jet definition}
\label{sec-kin}
A three-level trigger was used to select events online~\cite{zeus:1993:bluebook}.
The neutral current DIS events were selected offline using criteria similar to those
reported previously~\cite{hep-ex-0208037}. The main steps are outlined below.

The scattered electron was identified using an algorithm based on
a neural network~\cite{nim:a365:508,*nim:a391:360}.
The kinematic variables $Q^2$, $x_{{\rm Bj}}$ and the inelasticity $y$ were
reconstructed using the double-angle
method (DA)~\cite{proc:hera:1991:23,*proc:hera:1991:43}, 
where the hadronic final state was
reconstructed using combinations of CTD tracks and energy clusters measured
in the CAL to form energy-flow objects (EFOs)~\cite{epj:c1:81}.
% and $y$ using
%$e\Sigma$ method~\cite{nim:a426:583}
% epj:c6:43,briskin:phd:1998

The following criteria were applied offline to select DIS events:
\begin{itemize}
\item a scattered electron with energy $E_{e}^{'}$ above 10 GeV, to 
      ensure a well reconstructed electron and to suppress the 
      background from photoproduction events, in which the scattered 
      electron escapes undetected in the rear beampipe;
\item 40 $< \delta <$ 65 GeV, where $\delta = \sum_i (E_i-P_{{\rm Z},i})$,
      where $E_i$ and $P_{{\rm Z},i}$ are the energy and Z-component of the momentum
      of each EFO and the scattered electron. This cut removed events with large initial-state
     radiation and reduced the background from photoproduction events;
\item $y_{e}<0.95$, where 
      $y_{e}=1-\frac{E_{e}^{'}}{2E_{e}}(1-{\rm cos}~\theta_{e}^{'})$ and  
      $\theta_{e}^{'}$ is the polar angle of the electron. Along with the 
       previous requirements, this reduces the photoproduction background 
      to a negligible level;
\item $|X| > 24$ cm or $|Y| > 12$ cm, where
      $X$ and $Y$ are the impact positions of the
      positron on the CAL, to avoid the low-acceptance region adjacent to
      the rear beampipe;
\item the $Z$ coordinate of the vertex, $Z_{\rm vtx}$, determined 
      from CTD tracks, was required to be 
      in the range $\left|Z_{\rm vtx}\right| < 50~\rm{cm}$ along the beam axis.
      This cut removed background events from non-$ep$ interactions; 
\item $0.04<y_{\rm DA}<0.7$;
\item $20<Q^2_{\rm DA}<100\gev^2$;
\item $0.0004<x_{\rm DA}<0.005$.
\end{itemize}

After this selection, the jets were identified using the $k_T$ cluster 
algorithm~\cite{np:b406:187} in the longitudinally invariant inclusive 
mode~\cite{pr:d48:3160} applied in the Breit frame~\cite{zpf:c2:237}
on the CAL and FPC cells, excluding those belonging to the scattered 
electron. The reconstructed jets were then boosted back to the laboratory
frame. Jet-energy corrections were applied in order to account 
for the energy loss in the inactive material in front of the detector. 
The events were required to have at least one jet satisfying the following 
criteria in the laboratory frame:
\begin{itemize}
\item the transverse energy of each jet was required to be
     $E_{T}^{\rm jet}> 5~\mbox{GeV}$;
\item the pseudorapidity of each jet was required to be in the interval 
     $2 < {\eta}^{\rm jet} <4.3$;
\item $x^{\rm jet} > 0.036$, which selects forward jets with large energy;
\item $0.5 < (E_{T}^{\rm jet})^2/Q^{2} < 2$, which suppresses the DGLAP-type evolution;
\item jets with $2.8 < {\eta}^{\rm jet} <3.35$ and with the azimuthal angle of
the jet, ${\phi}$ expressed in radians, in the ranges
      $0 < {\phi}^{\rm jet} < 0.4$, $1.0 < {\phi}^{\rm jet} <2.2$,
      $2.7 < {\phi}^{\rm jet} <3.6$, $4.2 < {\phi}^{\rm jet} < 5.3$ or
      $5.7 < {\phi}^{\rm jet} <6.3$ were rejected due to poor reconstruction
      caused by the large cell size in the FCAL;
\end{itemize}

Using the sample described above, the triple differential cross sections, with the
$0.5 < (E_{T}^{\rm jet})^2/Q^{2} < 2$ cut removed, were measured in two intervals
of $Q^2$, $20<Q^2<40$ GeV$^2$ and $40<Q^2<100$ GeV$^2$ and for 
$E_{T}^{\rm jet}>5~\mbox{GeV}$.

For the ``forward jet+dijet'' analysis, the events were required to have one forward jet,
satisfying the same selection criteria as above, with the exception of the
$(E_{T}^{\rm jet})^2/Q^{2}$
cut, and at least two additional jets with $E_{T}^{\rm jet}> 5~\mbox{GeV}$.
The two additional jets, chosen with the highest transverse energy were required to
lie in the pseudorapidity region $-1.5 < {\eta}^{\rm jet} <4.3$.
The three selected jets were ordered in
${\eta}^{\rm jet}$ as described in Section 2.

% ---------------------------------------------------------------------------- %
%Section 5   MC Models and QCD Calculations% 
%----------------------------------------------------------------------------%

\section{Monte Carlo simulations} 
\label{sec-MC} 

Various MC samples were generated. Samples of {\sc Lepto} and {\sc Ariadne} generators were used
to simulate the detector response to jets of hadrons and to determine the hadronisation corrections
needed for comparison with perturbative QCD calculations. The same samples were used to
compare the measurements to the {\sc Lepto} and {\sc Ariadne} models. In addition,
the data were compared to expectations of the {\sc Ariadne} generator
with its newly tuned proton-remnant treatment, ``{\sc Ariadne} tuned'', and
to the {\sc Cascade} generator.

The {\sc Lepto}~\cite{cpc:101:108} MC program is based on first-order QCD
matrix elements supplemented with parton showers (MEPS), which follow DGLAP
evolution.

The CDM approach is represented by the {\sc Ariadne} 4.08~\cite{cpc:71:15} MC program
and its tuned variant (``{\sc Ariadne} tuned''\footnote{ The
following parameters have been changed from their default values: the powers, defining
the fraction of the proton remnant participating in the emission, PARA(10)=1.2
(default=1.0) and PARA(25)=1.2 (default=2.0); the square root of the mean value of the
primordial $p_{\perp}^{2}$ in the proton remnant, PARA(27)=0.9 (default=0.6).}) as
described in a recent H1 publication~\cite{desy-05-135}.

The {\sc Cascade} MC program~\cite{epj:c19:351,cpc:143:100}
is based on the CCFM evolution and uses $k_T$-factorisation of the cross section into an
off-shell matrix element and an unintegrated parton (gluon) density function
(uPDF). Predictions of {\sc Cascade} were obtained with the J2003 set-1 and set-2
uPDFs~\cite{spb:2004ugd}. The J2003 set-2 includes non-singular terms in the
splitting function and reduces the cross sections at low $x_{{\rm Bj}}$.  

In all MC models, the fragmentation of the final-state partons has been performed
using the {\sc Lund}~\cite{prep:97:31} string model as implemented in {\sc
Jetset} 7.4~\cite{cpc:82:74}.  

Both {\sc Lepto} and {\sc Ariadne} were interfaced to {\sc Heracles}~4.6.1~\cite{cpc:69:155}
via {\sc Djangoh}~1.1~\cite{cpc:81:381}. The {\sc Heracles}
program simulates first-order electroweak radiative corrections. The
CTEQ5L~\cite{pr:d55:1280} proton parton distribution functions (PDF) were used in
both cases.  

The events generated with {\sc Lepto} and {\sc Ariadne} were
passed through the {\sc Geant}~3.13-based~\cite{tech:cern-dd-ee-84-1} ZEUS
detector and trigger simulation programs~\cite{zeus:1993:bluebook}.
They were reconstructed and analysed using the same program chain as the
data.

% ----------------------------------------------------------------------------
\section{Acceptance correction and systematic studies}
\label{sec-accept}

The correction factors to the data for detector-acceptance effects were obtained
with the {\sc Ariadne} and {\sc Lepto} MC programs. These correction factors were
calculated bin by bin as
\begin{equation}
C_{\rm Acc} = \frac{N_{\rm MC}^{\rm det}}{N_{\rm MC}^{\rm had}},
\nonumber
\end{equation}

where $N_{\rm MC}^{\rm det}$ ($N_{\rm MC}^{\rm had}$) is the number of jets in bins of
the detector (hadron) level distribution. For this approach to be valid, the
distributions in the data must be well described by the MC simulation at the detector
level, a condition which was in general satisfied by both {\sc Ariadne} and {\sc Lepto}.
The average between the correction factors obtained with {\sc Ariadne} and {\sc Lepto}
was taken. The values of $C_{\rm Acc}$ were generally between 0.4 and 1.2 for the
inclusive forward-jet sample, and 0.5 to 1.4 for the ``forward jet+dijet'' sample.

To ensure the correct MC reconstruction of the jets near the boundary between FCAL
and FPC, jet profiles in the data and MC were compared and found to be in good
agreement.

The major sources of systematic uncertainty were as follows 
(the effects on the cross sections are shown in parentheses):

\begin{itemize}
\item the largest uncertainty resulted from the model dependence of the
      acceptance corrections. This uncertainty was estimated using the deviations
      of {\sc Lepto} and {\sc Ariadne} corrections from their average 
      ($\le 10$\% for inclusive forward jet sample, $<25~$\% for forward jet+dijet sample);
\item $\pm~3$\% shift of jet energies due to the CAL energy-scale uncertainty ($<10~$\% );
\item $\pm~10$\%  shift of jet energies due to the FPC energy-scale uncertainty
      ($\sim15$\% for the last $\eta^{\rm jet}$ bin, negligible elsewhere);
\item the selection of inclusive DIS events ($<1$\%).
      The cuts on the scattered-electron energy, the $X$ and $Y$ position of the
      electron, $\delta$ and $Z_{\rm vtx}$ were varied.
\end{itemize}

These systematic uncertainties, except for the energy-scale uncertainty, were
added in quadrature separately for the positive and negative variations in each
bin. The energy scale uncertainties, which are correlated between bins, are
shown separately. The uncertainty in the luminosity of $\pm~2.2$\% is not included
in the figures.  

% ---------------------------------------------------------------------------- %
%Section 7 
%----------------------------------------------------------------------------%
\section{NLO QCD calculations} 
\label{sec-QCD} 

For inclusive forward jets, fixed-order calculations were performed with the
{\sc Disent}~\cite{np:b485:291} code, at ${\cal O}(\alpha \alpha_{s}^{2})$ in the
$\overline{\rm MS}$ renormalisation and factorisation schemes. The number of
flavours was set to five; the renormalisation ($\mu_{\rm R}$) and factorisation
($\mu_{\rm F}$) scales were both set to $\mu_{\rm R} = \mu_{\rm F} =Q$. The
CTEQ6M~\cite{jhep:0602:032} parameterisation of the proton PDFs was used. The
theoretical uncertainty in the calculations was estimated considering the following three
sources:
\begin{itemize}
\item $\mu_{\rm R}$ was varied up and down by a factor of two, which
contributed up to $57$\% depending on the phase-space region;
\item $\mu_{\rm F}$ was also varied up and down by a factor of two and the resulting
uncertainty was less than $5$\% except in the lowest $x_{\rm Bj}$ bin and the
most forward region;
\item the PDF uncertainty was estimated using the 40 different sets
of CTEQ6 parton distribution functions~\cite{jhep:0602:032}.
\end{itemize}

For ``forward jet+dijet'' cross sections, the program {\sc Nlojet++}~\cite{nlojet} was used.
This program calculates three-jet production in DIS at NLO 
${\cal O}({\alpha}{\alpha}_s^3)$ and uses the $\overline{\rm MS}$ 
scheme. The number of flavours, $\mu_{\rm R}$
and $\mu_{\rm F}$ were set as in {\sc Disent}.

In order to compare the data to NLO calculations, corrections from the parton to the
hadron level, $C_{HAD}$, were determined in each bin. The hadronisation corrections 
$C_{HAD}$ were calculated as an average between those obtained from {\sc Lepto} and from
{\sc Ariadne} and applied to the NLO calculations. The uncertainty of the hadronisation
correction $C_{HAD}$ was assumed to be the absolute difference in the two values.

% ---------------------------------------------------------------------------- %
%Section 8
%----------------------------------------------------------------------------
\section{Results} \label{results}

Cross sections for inclusive forward jets and for events containing a dijet system
in addition to the forward jet, were measured in the kinematic region given by
$20<Q^2\!<\!100\gev^2$, $0.04<y<0.7$ and $0.0004 < x_{\rm Bj} < 0.005$.  The
differential jet cross sections are presented as functions of the variable 
$\xi = Q^2, x_{\rm Bj}, E_{T}^{\rm jet}, {\eta}^{\rm jet}$. They were determined
as

\begin{equation}
\frac{d\sigma}{d\xi} = \frac{N^{\rm jet}_{\data}}
{\mathcal{L}\cdot\Delta \xi}\cdot 
  \frac{C^{\rm QED}}{C_{\rm Acc}},
\nonumber
\end{equation}
where $N^{\rm jet}_{\data}$ is the number of jets in a bin of width $\Delta \xi$,
$\mathcal{L}$ is the integrated luminosity, and $C^{\rm QED}=N_{\rm MC}^{\rm no
QED}/N_{\rm MC}^{\rm QED}$, where $N_{\rm MC}^{\rm QED}$ ($N_{\rm MC}^{\rm no QED}$) is
the number of events selected at the hadron level in a given $\xi$ bin in the MC sample
generated with (without) QED radiation.

The triple differential cross sections were obtained in a similar manner,

\begin{equation}
\frac{d^3\sigma}{dQ^2 d(E_{T}^{\rm jet})^2 d\eta} = \frac{N^{\rm jet}_{\rm data}}
{\mathcal{L}\cdot\Delta Q^2 \cdot\Delta (E_{T}^{\rm jet})^2 \cdot \Delta \eta }\cdot
\frac{C^{\rm QED}}{C_{\rm Acc} }.
\nonumber
\label{equ:tripldiff}
\end{equation} 

% ----------------------------------------------------------------------------
\subsection{Inclusive forward-jet measurements}
\label{results-fjet}

\begin{figure}[tbhp]
\begin{center}
\includegraphics[height=100mm]{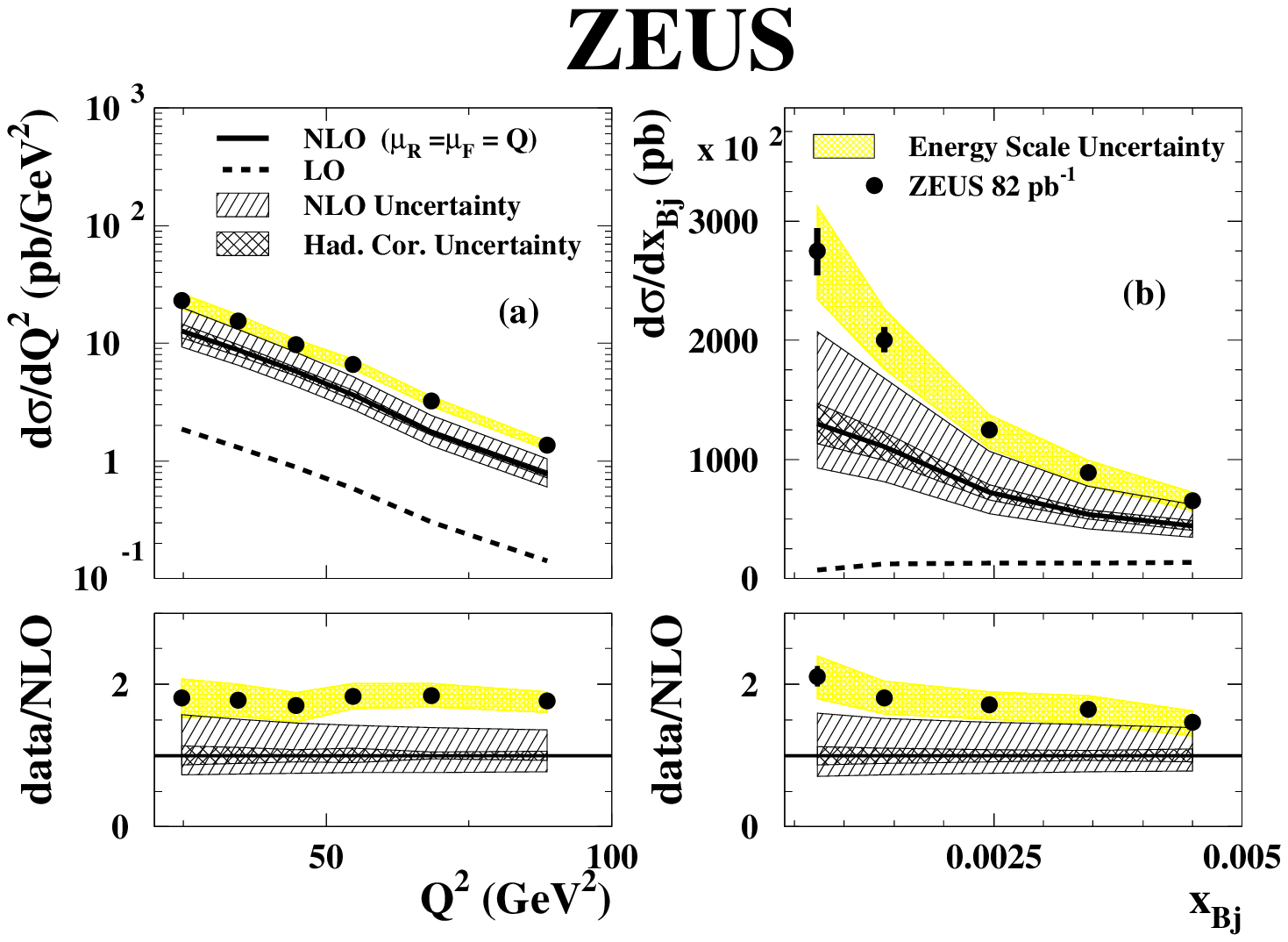}
\includegraphics[height=90mm]{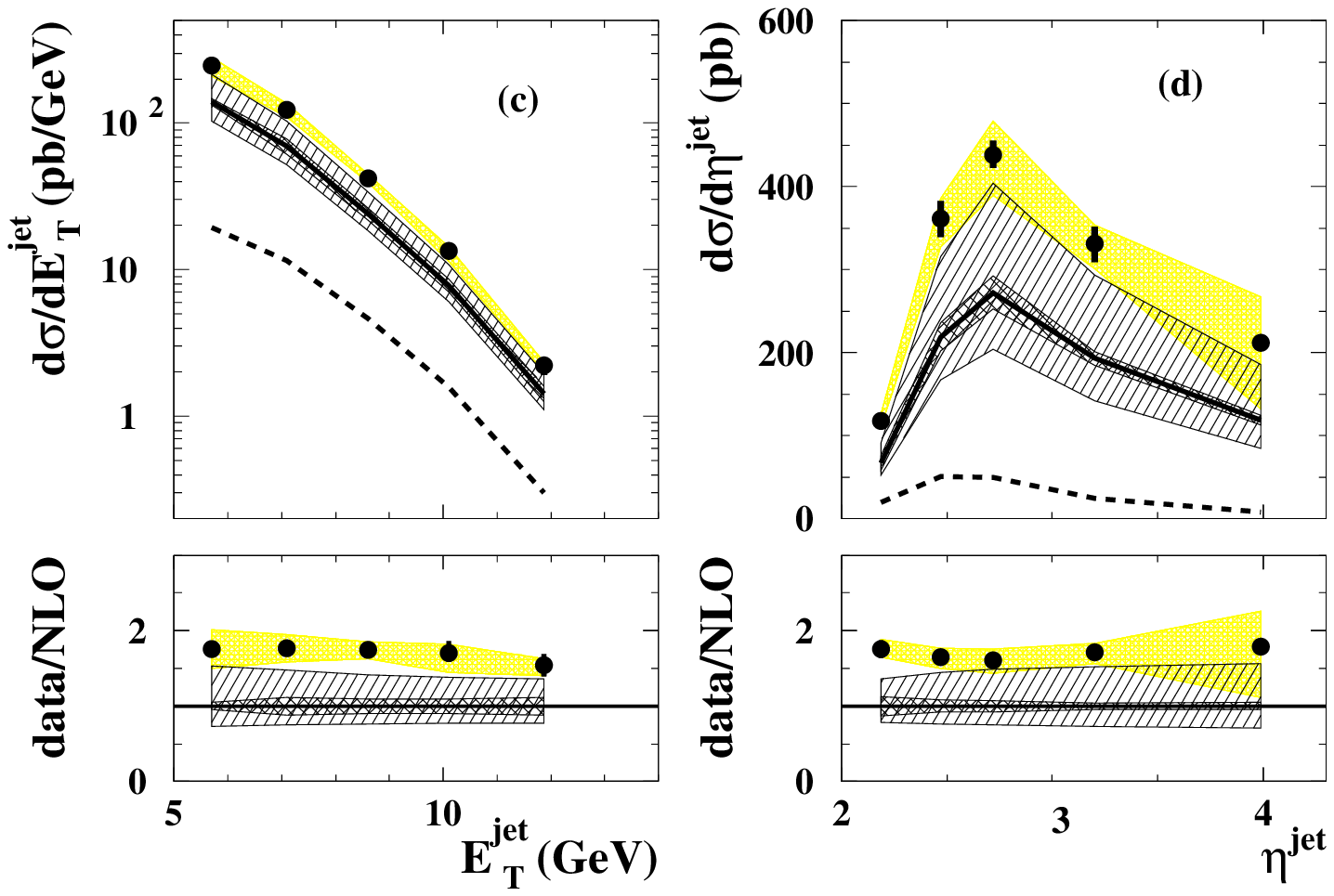}
\end{center}
\caption{Measured differential cross sections as a function of 
(a) $Q^2$, (b) $x_{\rm Bj}$, (c) $E_{T}^{jet}$ and (d) $\eta^{jet}$
for inclusive jet production (dots) compared with the NLO QCD calculations (solid
line). The hatched area shows the theoretical uncertainties and
the shaded area shows the uncertainty after
varying the CAL and FPC energy scales. The inner error bars indicate the
statistical uncertainties, while the outer ones correspond to
statistical and systematic uncertainties (except the energy-scale
uncertainty) added in quadrature.
}\label{fig-fig1}
\end{figure}

The measured differential forward-jet cross sections as functions of $Q^2$,
$x_{\rm Bj}$, $E_{T}^{\rm jet}$ and $\eta^{\rm jet}$ are shown in
Fig.~\ref{fig-fig1}, where they are compared to NLO calculations.  

The calculations predict lower cross sections than obtained from the data
by as much as a factor two; however, they have a
large theoretical uncertainty. The strong dependence of the calculation on
$\mu_{\rm R}$ can be related to the fact that in this kinematic region
higher-order terms become relevant. As a demonstration, the leading-order
calculation is also shown in Fig.~\ref{fig-fig1} for each differential cross
section. It is far below the measurement, indicating that the contribution of
${\cal O}(\alpha_{\rm s}^{2})$ terms is significant. A
recent publication~\cite{desy-05-135}, which used a harder renormalisation scale (the average
$E_T^2$ of the dijets coming from the hard scattering), reported a smaller
renormalisation-scale uncertainty.

\begin{figure}[tbhp]
\begin{center}
\includegraphics[height=160mm]{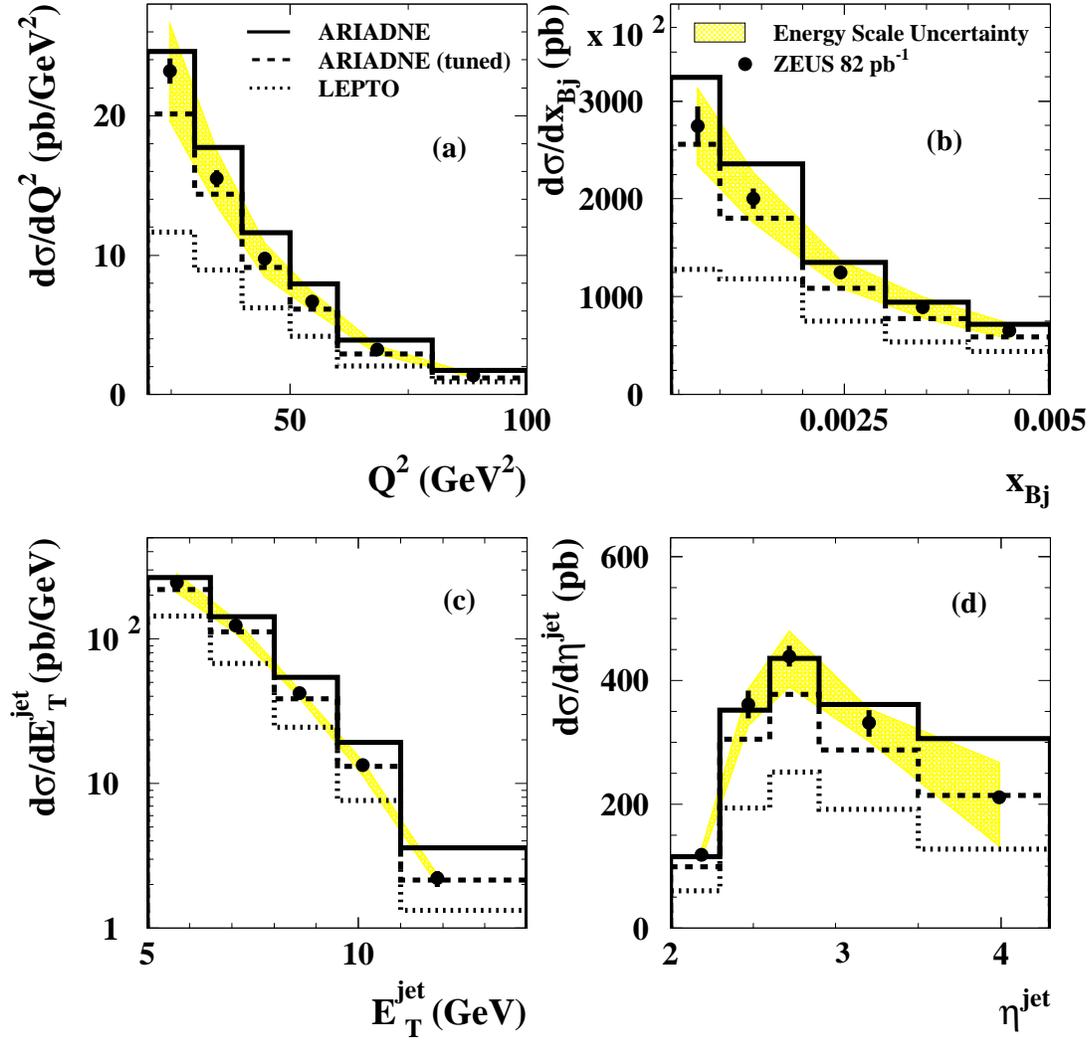}
\end{center}
\caption{Measured differential cross sections as a function of
(a) $Q^2$, (b) $x_{\rm Bj}$, (c) $E_{T}^{jet}$ and (d) $\eta^{jet}$
for inclusive jet production (dots) compared with the
ARIADNE (solid histogram), ARIADNE with new tuning (dashed histogram) 
and LEPTO (dotted histogram) predictions.
The shaded area shows the uncertainty after
varying the CAL and FPC energy scales.
The inner error bars indicate the
statistical uncertainties, while the outer ones correspond to
statistical and systematic uncertainties (except the
energy-scale uncertainty) added in quadrature.
}\label{fig-fig2}
\end{figure}

A comparison of the data with the {\sc Ariadne} and {\sc Lepto} MC is shown in
Fig.~\ref{fig-fig2}. The predictions of the CDM obtained with ``{\sc Ariadne} default'' are
in fair agreement with the data with the
exception of high $E_{T}^{\rm jet}$ and high $\eta^{\rm jet}$, where {\sc Ariadne}
overestimates the cross sections. An investigation has shown
that in {\sc Ariadne} the proton-remnant fragments are generated with high $p_T$,
therefore they show up at much lower $\eta$ than in other generators. The newly tuned
{\sc Ariadne}, also shown in Fig.~\ref{fig-fig2}, yields lower cross sections, in
particular at high $E_T^{\rm jet}$ and high $\eta^{\rm jet}$, and provides a good
description of the data.

The predictions of the {\sc Lepto} MC are found to be in agreement with data in shape
for all distributions, however the absolute normalisation is below the
measurements by a factor of two.
 
\begin{figure}[tbhp]
\begin{center}
\includegraphics[height=160mm]{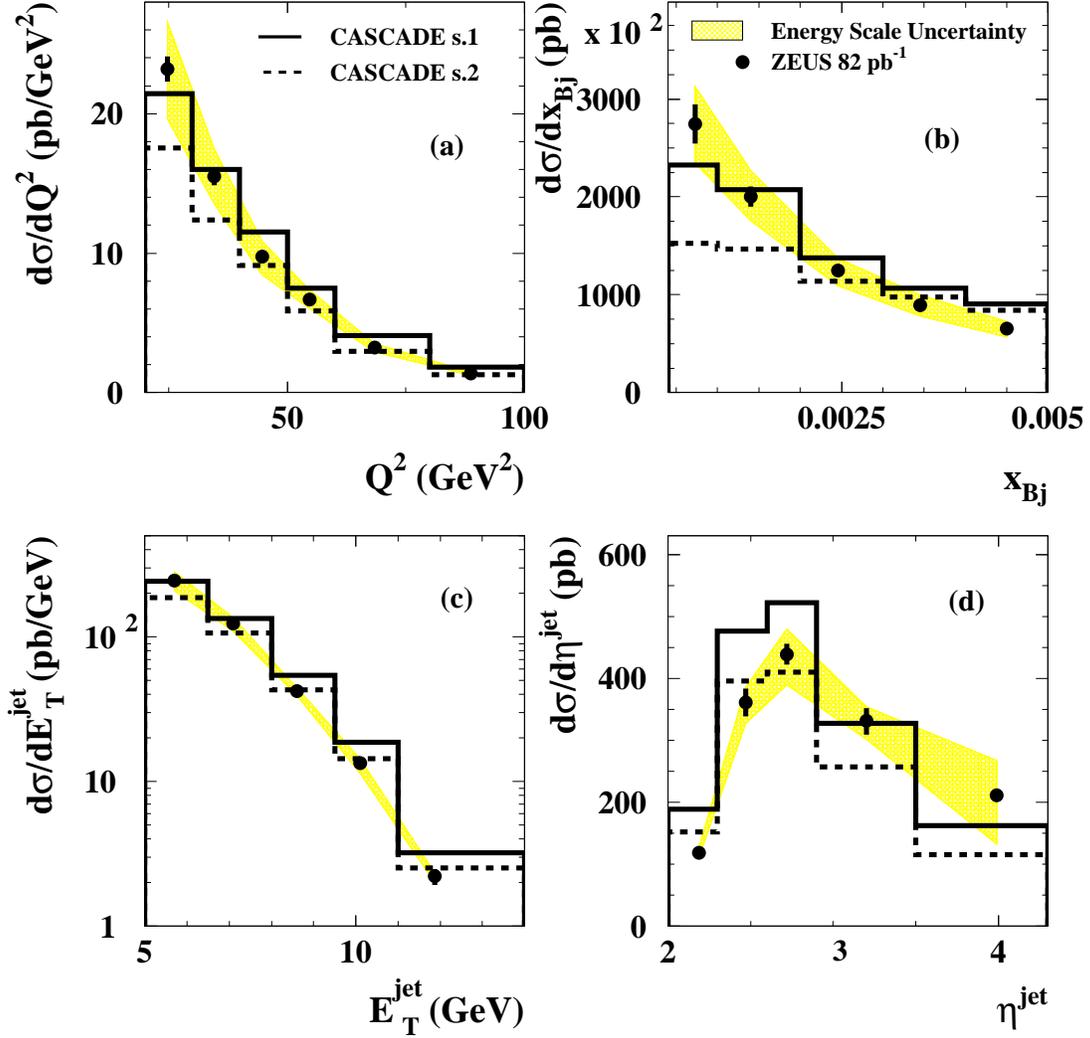}
\end{center}
\caption{Measured differential cross sections as a function of
(a) $Q^2$, (b) $x_{\rm Bj}$, (c) $E_{T}^{jet}$ and (d) $\eta^{jet}$
for inclusive jet production (dots) compared with the
CASCADE set-1 parametrisation (solid histogram) and 
CASCADE set-2 (dashed histogram) predictions.
The shaded area shows the uncertainty after
varying the CAL and FPC energy scales.
The inner error bars indicate the
statistical uncertainties, while the outer ones correspond to
statistical and systematic uncertainties (except the
energy-scale uncertainty) added in quadrature.
}\label{fig-fig3}
\end{figure}

The measurement of differential forward-jet cross sections is compared to the prediction
of the {\sc Cascade} MC model in Fig.~\ref{fig-fig3}. Neither of the investigated
uPDF sets gives a satisfactory agreement with the measurements in all distributions,
suggesting that a further adjustment of the input parameters of the {\sc Cascade} model is
necessary.

%----------------------------------------------------------------------------
\subsection{Triple-differential forward-jet cross section} \label{results-fjet-triple}

\begin{figure}[tbhp]
\begin{center}
\includegraphics[height=160mm]{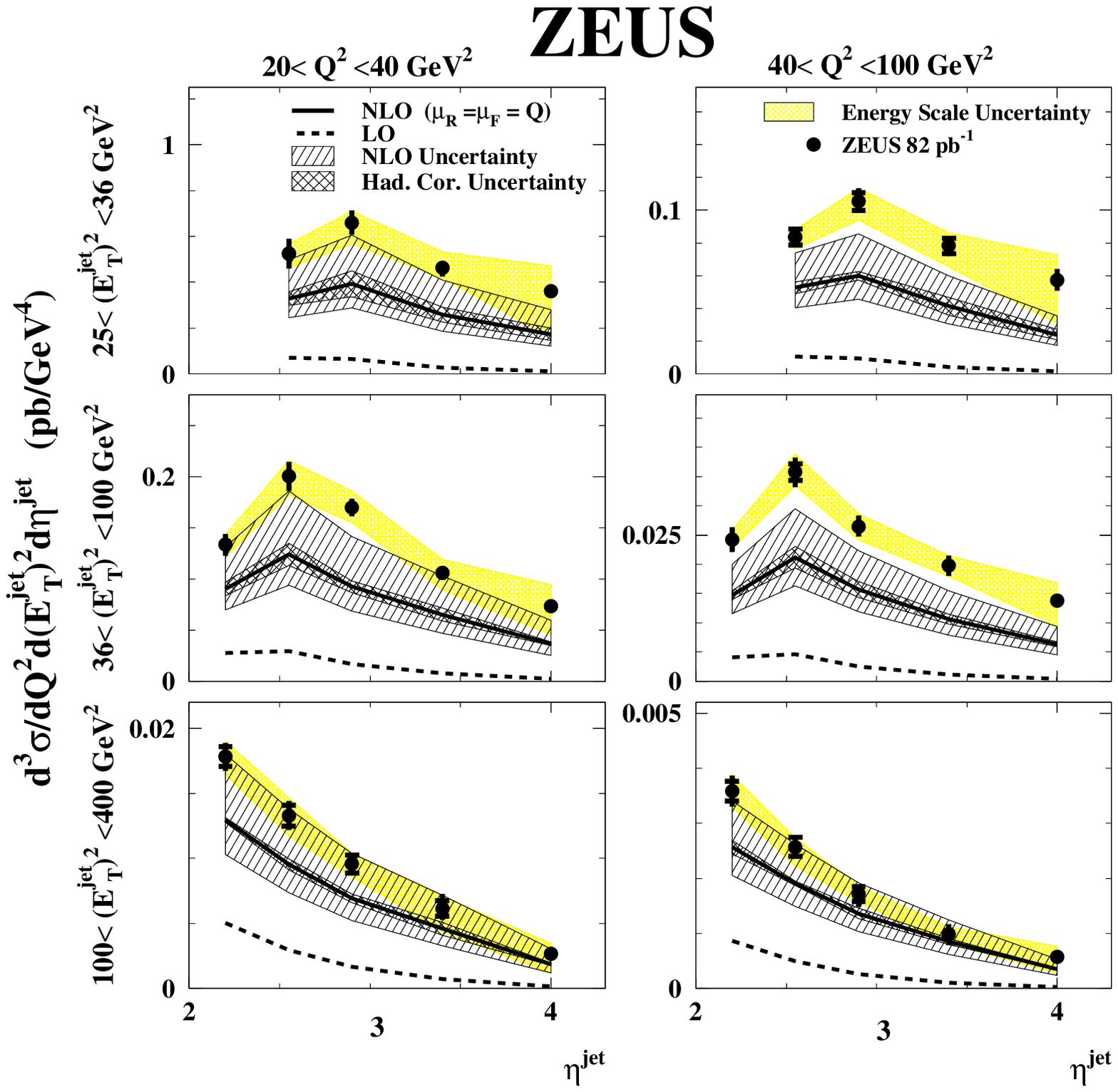}
\end{center}
\caption{Measured differential cross sections as a function of
$\eta^{jet}$ in different bins of $Q^2$ and $(E_{T}^{jet})^2$
for inclusive jet production (dots) compared with the NLO QCD calculations (solid
line). The hatched area shows the theoretical uncertainties and
the shaded area shows the uncertainty after
varying the CAL and FPC energy scales.
The inner error bars indicate the
statistical uncertainties, while the outer ones correspond to
statistical and systematic uncertainties (except the
energy-scale uncertainty) added in quadrature.
}\label{fig-triple-nlo}
\end{figure}

The triple-differential forward-jet cross sections as a function of $\eta^{\rm jet}$ are
presented in two intervals of $Q^{2}$ and three intervals of $(E_{T}^{\rm jet})^2$ in
Fig.~\ref{fig-triple-nlo}. Also shown in Fig.~\ref{fig-triple-nlo} are the expectations
of the NLO calculations from {\sc Disent}. The calculations generally underestimate the
cross sections. The largest discrepancy between the data and the theory is seen in the
high-$Q^{2}$ range and for $(E_{T}^{\rm jet})^2 < 100~\gev^2$. This region is sensitive to
multigluon emission, which is lacking in the NLO calculations.

\begin{figure}[tbhp]
\begin{center}
\includegraphics[height=160mm]{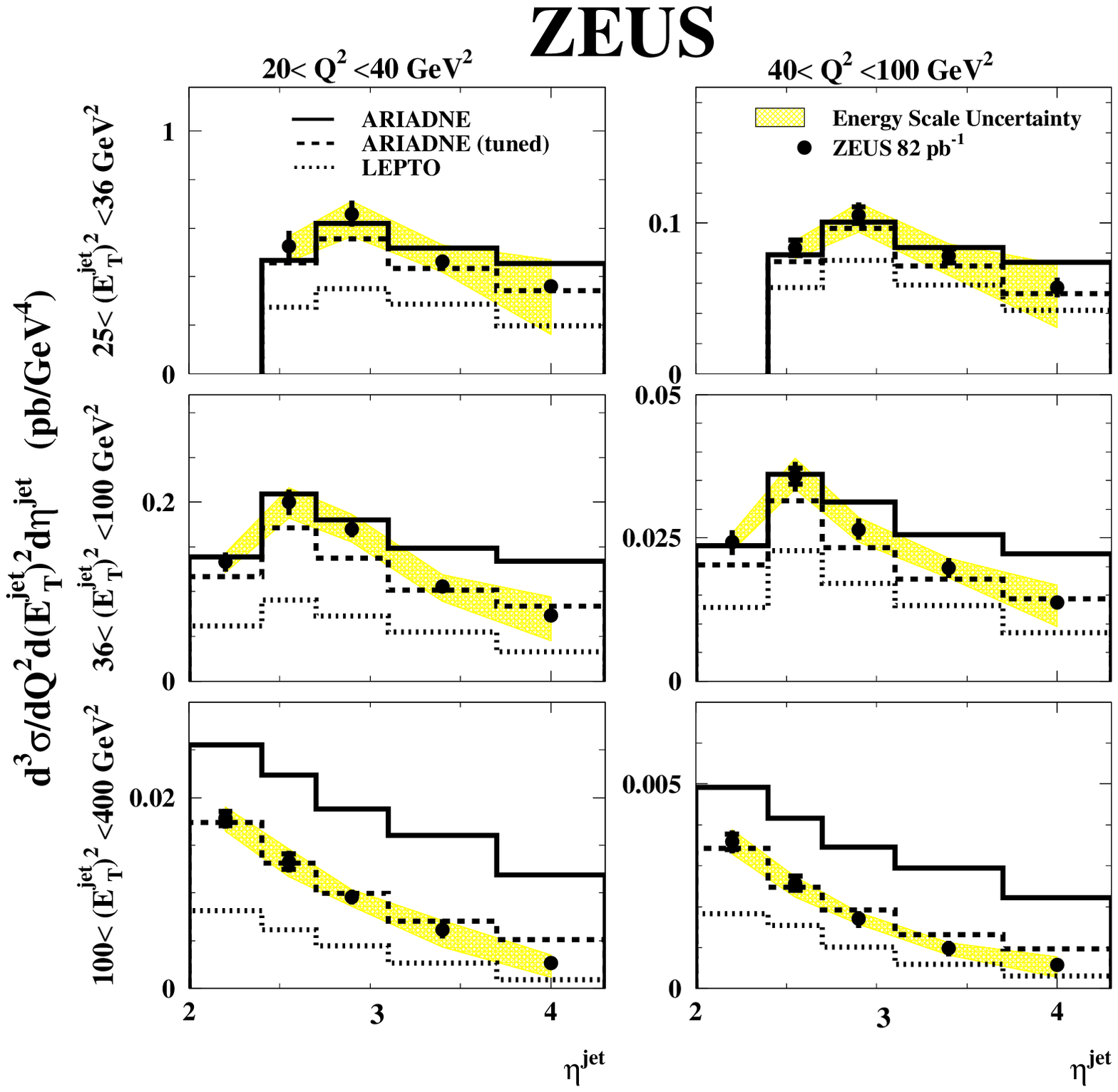}
\end{center}
\caption{Measured differential cross sections as a function of
$\eta^{jet}$ in different bins of $Q^2$ and $(E_{T}^{jet})^2$
for inclusive jet production (dots) compared with the
ARIADNE (solid histogram), ARIADNE with new tuning (dashed histogram) 
and LEPTO (dotted histogram) predictions.
The shaded area shows the uncertainty after
varying the CAL and FPC energy scales.
The inner error bars indicate the
statistical uncertainties, while the outer ones correspond to
statistical and systematic uncertainties (except the
energy-scale uncertainty) added in quadrature.
}\label{fig-triple-al}
\end{figure}

In Fig.~\ref{fig-triple-al}, the data are compared with {\sc Lepto} and {\sc
Ariadne}. As was already observed in Fig.~\ref{fig-fig2}, the {\sc Lepto} MC is always
below the measurements.

The cross sections of ``{\sc Ariadne} tuned'' are below those of the ``{\sc Ariadne} default''
in all the presented phase space. The difference between the two versions is smallest
in the lowest-$(E_{T}^{\rm jet})^2$ interval, where both are close to the data. In
the highest-$(E_{T}^{\rm jet})^2$ interval, the difference is big and ``{\sc Ariadne} tuned''
gives a good description of the data.

\begin{figure}[tbhp]
\begin{center}
\includegraphics[height=160mm]{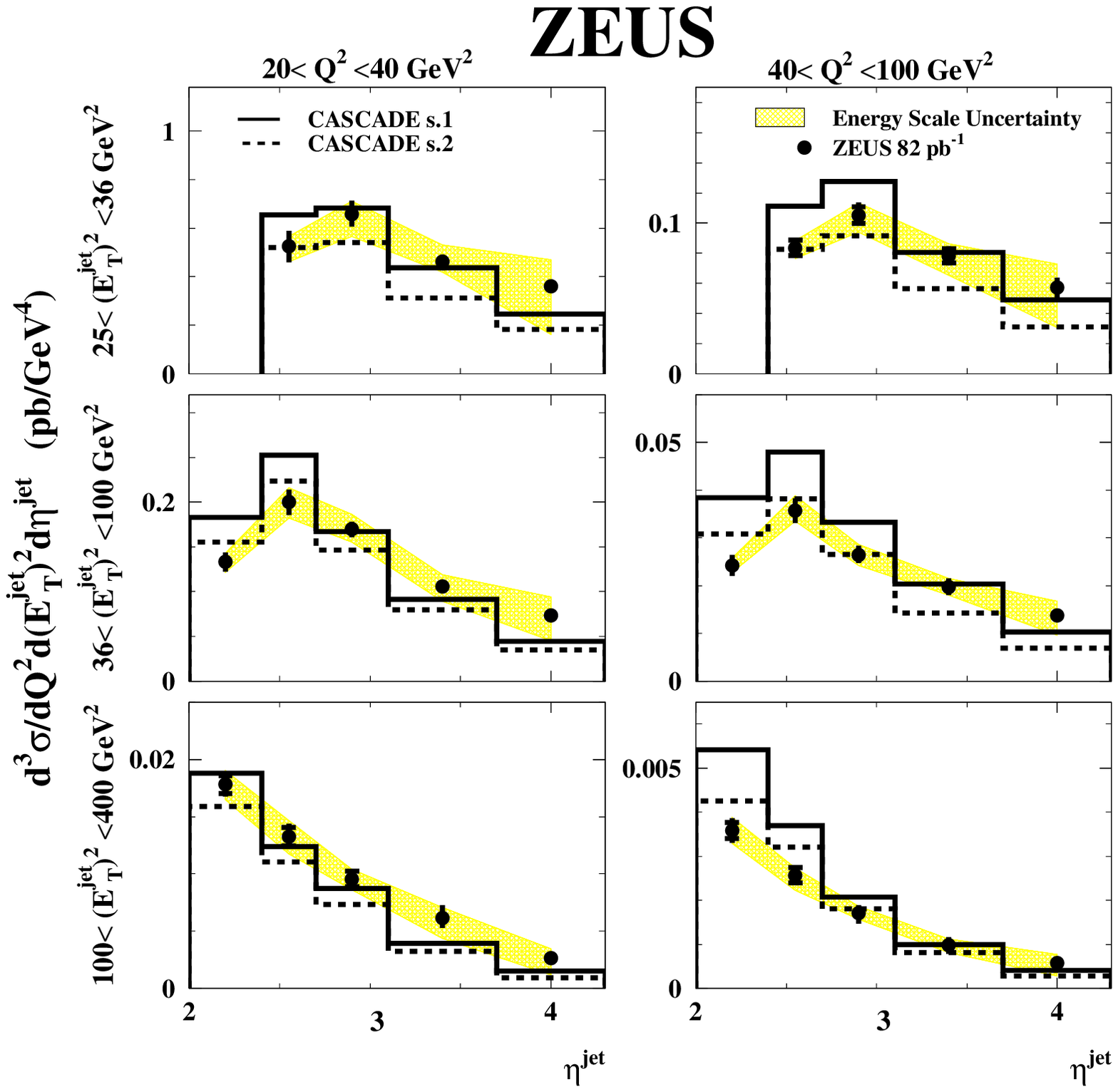}
\end{center}
\caption{Measured differential cross sections as a function of
$\eta^{jet}$ in different bins of $Q^2$ and $(E_{T}^{jet})^2$
for inclusive jet production (dots) compared with the
CASCADE set-1 parametrisation (solid histogram) and 
CASCADE set-2 (dashed histogram) predictions.
The shaded area shows the uncertainty after
varying the CAL and FPC energy scales.
The inner error bars indicate the
statistical uncertainties, while the outer ones correspond to
statistical and systematic uncertainties (except the
energy-scale uncertainty) added in quadrature.
}\label{fig-triple-ca}
\end{figure}

A comparison of the data with the {\sc Cascade} MC with two sets of uPDFs
is shown in Fig.~\ref{fig-triple-ca}. 

The expectations of {\sc Cascade} are close to the data in the low-$Q^{2}$
interval for set-1, while in the high-$Q^{2}$ interval the set-2 gives a
better description of the data. None of the sets can accomodate all the features of the data.

% ----------------------------------------------------------------------------
\subsection{Forward jet+dijet measurements}
\label{results-fjet-dijet}

\begin{figure}[tbhp]
\begin{center}
\includegraphics[height=160mm]{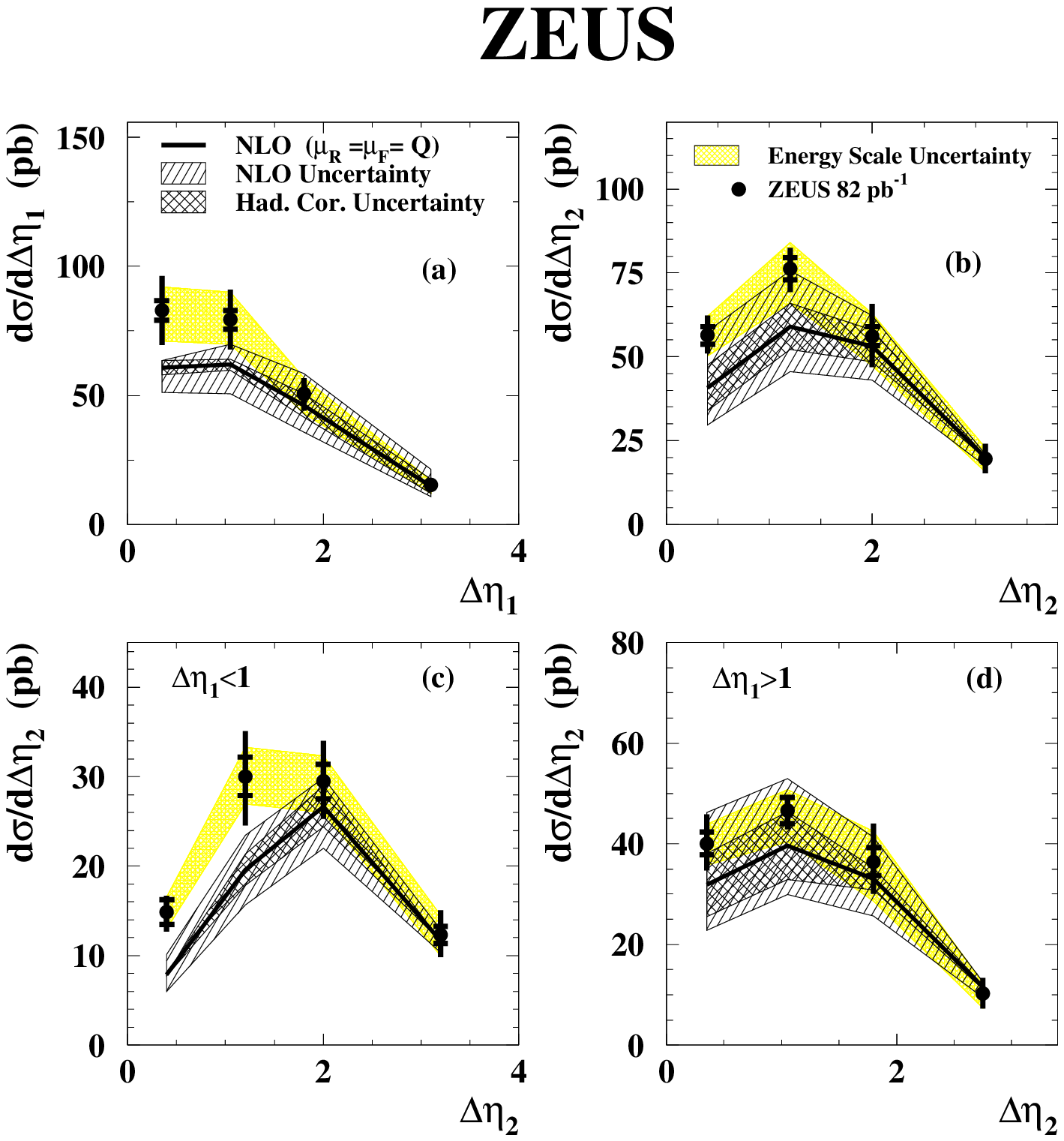}
\end{center}
\caption{ Differential cross sections for forward+dijet sample
as a function of (a) $\Delta\eta_{1}$, (b) $\Delta\eta_{2}$,
(c) $\Delta\eta_{2}$ for $\Delta\eta_{1}<1$ and (d) $\Delta\eta_{2}$ for
$\Delta\eta_{1}>1$.
The data (dots) are compared with the NLO QCD calculations (solid
line). The hatched area shows the theoretical uncertainties.
The shaded area shows the uncertainty after varying the CAL and FPC energy 
scales. The inner error bars indicate the statistical uncertainties, 
while the outer ones correspond to statistical and systematic uncertainties 
(except the energy-scale uncertainty) added in quadrature.}
\label{fig-dijetnlo}
\end{figure}

\begin{figure}[tbhp]
\begin{center}
\includegraphics[height=160mm]{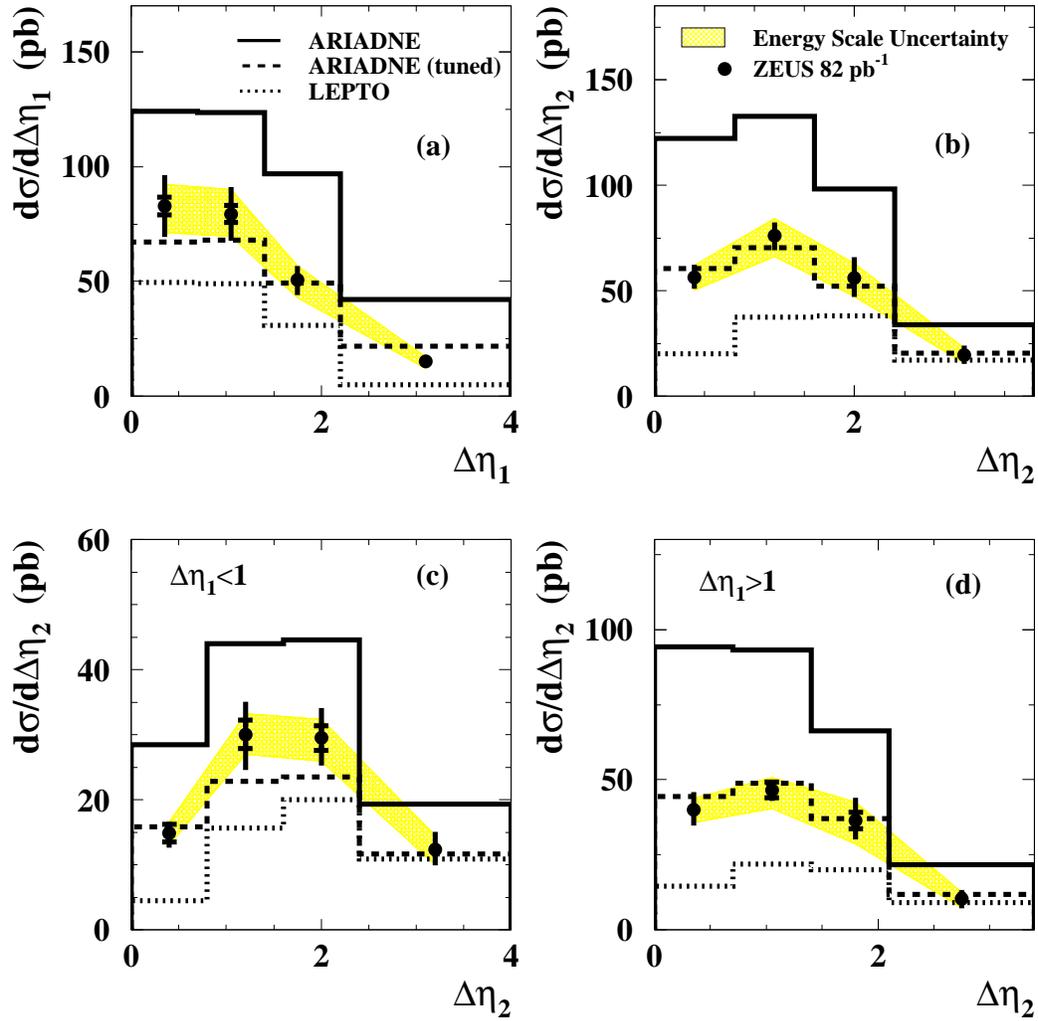}
\end{center}
\caption{ The differential cross sections for forward+dijet sample
as a function of (a) $\Delta\eta_{1}$, (b) $\Delta\eta_{2}$,
(c) $\Delta\eta_{2}$, for $\Delta\eta_{1}<1$ and (d) $\Delta\eta_{2}$ for
$\Delta\eta_{1}>1$.
The data (dots) are compared with the ARIADNE (solid histogram), 
ARIADNE with new tuning (dashed histogram) and LEPTO
(dotted histogram) predictions. The shaded area shows the
uncertainty after varying the CAL and FPC energy scales.
The inner error bars indicate the statistical uncertainties,
 while the outer ones correspond to
statistical and systematic uncertainties (except the energy-scale
uncertainty) added in quadrature.}
\label{fig-fig4}
\end{figure}

\begin{figure}[tbhp]
\begin{center}
\includegraphics[height=160mm]{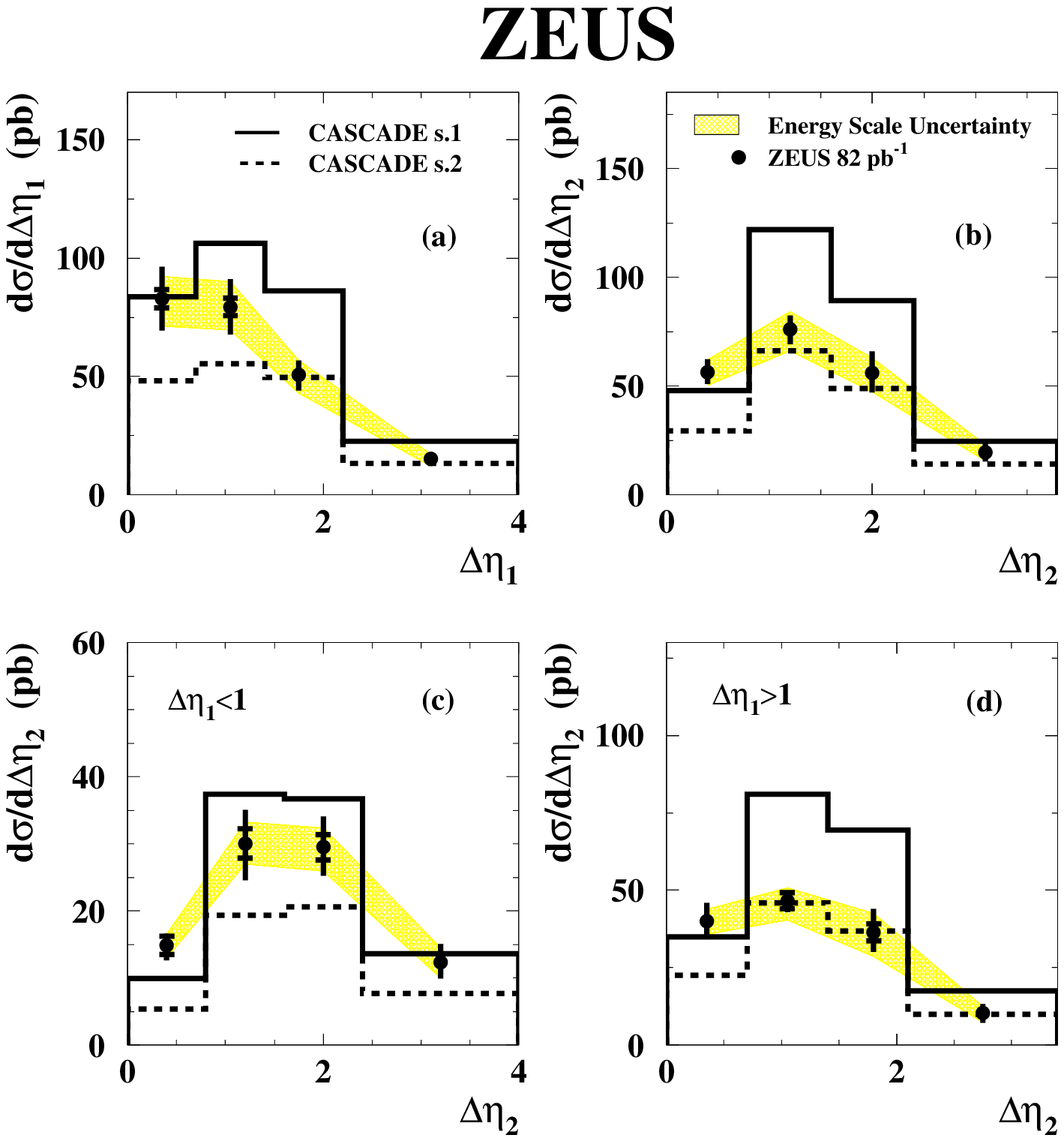}
\end{center}
\caption{ The differential cross sections for forward+dijet sample
as a function of (a) $\Delta\eta_{1}$, (b) $\Delta\eta_{2}$,
(c) $\Delta\eta_{2}$, for $\Delta\eta_{1}<1$ and (d) $\Delta\eta_{2}$ for
$\Delta\eta_{1}>1$.
The data (dots) are compared with the CASCADE set-1 (solid histogram) and CASCADE set-2
(dashed histograms) predictions. The shaded area shows the
uncertainty after varying the CAL and FPC energy scale.
The inner error bars indicate the statistical uncertainties,
 while the outer ones correspond to
statistical and systematic uncertainties (except the energy-scale
uncertainty) added in quadrature.}
\label{fig-fig5}
\end{figure}

The measured cross sections for events with a dijet system in addition to the  forward
jet are compared with fixed-order QCD calculations and LO parton-shower MC models in
Figs.~\ref{fig-dijetnlo}-\ref{fig-fig5}. The cross sections are presented as a
function of $\Delta{\eta}_1$ and $\Delta{\eta}_2$, and as a function of $\Delta{\eta}_2$
for two cases, namely, $\Delta{\eta}_1 < 1$ and $\Delta{\eta}_1 > 1$.  

A comparison between data and the predictions of {\sc Nlojet++} is shown in
Fig.~\ref{fig-dijetnlo}. As already observed by the H1 experiment~\cite{desy-05-135},
the {\sc Nlojet++} calculations agree well with the data at large $\Delta{\eta}_2$,
while they do not describe the data at small $\Delta{\eta}_2$, especially for
small $\Delta{\eta}_1$. The large $\Delta{\eta}_2$ kinematics at low $x_{\rm Bj}$
favours dijets originating from photon-gluon fusion, with an additional gluon
responsible for the forward jet. This case is well treated by {\sc Nlojet++}. The small
$\Delta{\eta}_1$ and $\Delta{\eta}_2$ region corresponds to the event configuration
in which all the three jets tend to go forward, away from the hard interaction. This
configuration favours multigluon emission, which is not expected to be described by
{\sc Nlojet++}.

The comparison between data and the {\sc Lepto} and {\sc Ariadne} MCs is shown in
Fig.~\ref{fig-fig4}. As before, the {\sc Lepto} predictions are below the data for
all differential cross sections. The ``{\sc Ariadne} default'' overestimates the cross
sections. This implies that energetic multiple jets are produced too often in the
``{\sc Ariadne} default''. The tuning of the {\sc Ariadne} parameters brings this
model into very good agreement with data for all differential distributions.  

The comparison of the {\sc Cascade} MC to the data is presented in
Fig.~\ref{fig-fig5}. As before, {\sc Cascade} does not
satisfactorily reproduce the measurement.

% ----------------------------------------------------------------------------
\section{Summary}
\label{sec-summary}

A new measurement of the inclusive jet cross sections has been performed in an
extended forward region, $2 < \eta^{\rm jet} < 4.3$, with higher statistics and
smaller systematic uncertainties compared to previous studies.  The measured
differential cross sections are presented as functions of $Q^2$, $x_{\rm Bj}$,
$E_T^{\rm jet}$ and $\eta^{\rm jet}$. The measurements were compared to the
predictions of next-to-leading-order QCD calculations, which were found to be
below the data, in certain regions by as much as a factor of two. The large
contribution of next-to-leading-order corrections and the size of the theoretical
uncertainty indicate that in this phase space higher-order contributions are
important. The best overall description of the inclusive forward-jet cross
sections was obtained by the newly tuned {\sc Ariadne} MC model. The {\sc Cascade}
MC with J2003 set-1 and J2003 set-2 for 
unintegrated gluon density failed to satisfactorily
describe the data. Therefore, these measurements can be used for further
adjusting the input parameters of the {\sc Cascade} model.  

The measurement of the cross sections of the events containing a dijet system in
addition to the forward jet is presented as functions of pseudorapidity
separation between jets composing the dijet, $\Delta{\eta}_1$, and pseudorapidity
separation between forward jet and dijet system, $\Delta{\eta}_2$, for all
 $\Delta{\eta}_1$ values and for $\Delta{\eta}_1 < 1$ and $\Delta{\eta}_1 > 1$. NLO
calculations describe the data at large $\Delta{\eta}_2$ but underestimate
the cross sections at small $\Delta{\eta}_2$, especially for small values of
$\Delta{\eta}_1$, where, in the case of small $x_{\rm Bj}$, the contribution of
multiple gluon emission is expected to be large. The predictions of {\sc Lepto}
are significantly below the data. {\sc Ariadne} with default parameters
significantly overestimates the cross sections whereas the new tuning provides
a good description of the data. The {\sc Cascade} MC, as in the inclusive case, does
not provide a satisfactory description of measured cross sections.

%\newpage   
%------------------------------------------------------------------------
%       Acknowledgements
%------------------------------------------------------------------------
\vspace{2cm}
\noindent {\Large\bf Acknowledgments}
\vspace{1cm}

We thank the DESY Directorate for the strong support and encouragement.
We are grateful for the support of the DESY computing and network services.
The diligent efforts of the HERA machine group are gratefully acknowledged.
The design, construction and installation of the ZEUS detector have
been made possible due to the ingenuity and efforts of many people from 
DESY and other institutes who are not listed as authors. We also thank Hannes 
Jung and Herman Hessling for fruitful discussions.

\vfill\eject
\providecommand{\etal}{et al.\xspace}
\providecommand{\coll}{Coll.\xspace}
\catcode`\@=11
\def\@bibitem#1{%
\ifmc@bstsupport
  \mc@iftail{#1}%
    {;\newline\ignorespaces}%
    {\ifmc@first\else.\fi\orig@bibitem{#1}}
  \mc@firstfalse
\else
  \mc@iftail{#1}%
    {\ignorespaces}%
    {\orig@bibitem{#1}}%
\fi}%
\catcode`\@=12

%-------------------------------------------------------------------------------
%       An example table
%-------------------------------------------------------------------------------
\begin{table}[p]
\begin{center}
\begin{tabular}{||c|c|c|c|c|c||c||c||}
\hline
$Q^2$ bin & $d\sigma/dQ^2$ & $\delta_{stat}$ 
& $\delta_{syst}$ & $\delta_{CAL}$ & $\delta_{FPC}$ &\footnotesize $C_{QED}$&\footnotesize $C_{HAD}$\normalsize\\
$\rm{(GeV^2)}$ &$\rm{(pb/GeV^2)}$ & & & & & & \normalsize\\
 \hline\hline
20 - 30&  $23.21$  & $\pm 0.43$ & $^{+0.81}_{-0.79}$ & $^{+1.39}_{-1.95}$  & $^{+3.06}_{-3.03}$& $0.97$ & $0.89$ \\
30 - 40&  $15.49$  & $\pm 0.33$ & $^{+0.48}_{-0.54}$ & $^{+0.95}_{-1.05}$ & $^{+1.73}_{-1.61}$ & $0.96$ & $0.89$ \\
40 - 50&  $9.76 $  & $\pm 0.26$ & $^{+0.26}_{-0.37}$ & $^{+0.54}_{-0.77}$  & $^{+0.91}_{-1.07}$& $0.94$ & $0.89$ \\
50 - 60&  $6.65 $  & $\pm 0.21$ & $^{+0.27}_{-0.22}$ & $^{+0.40}_{-0.36}$ & $^{+0.55}_{-0.51}$ & $0.97$ & $0.91$ \\
60 - 80&  $3.21 $  & $\pm 0.10$ & $^{+0.12}_{-0.08}$ & $^{+0.20}_{-0.18}$ & $^{+0.22}_{-0.22}$ & $0.97$ & $0.89$ \\
80 - 100& $1.36 $  & $\pm 0.07$ & $^{+0.07}_{-0.03}$ & $^{+0.06}_{-0.08}$ & $^{+0.08}_{-0.09}$ & $0.94$ & $0.91$ \\
\hline
\end{tabular}
\caption{ The differential cross section, $d\sigma/dQ^2$, in bins of 
$Q^2$ for inclusive forward jets. The statistical ($\delta_{stat}$), systematic ($\delta_{syst}$) and 
jet-energy-scale uncertainties for CAL and FPC ($\delta_{CAL}$ and $\delta_{FPC}$) are shown separately.
The multiplicative correction applied to correct for QED radiative effects ($C_{QED}$) and for
hadronisation effects ($C_{HAD}$) are shown in the last two columns.}
\label{tab-1}
\end{center}
\end{table}

%%%%%%%%%%%%%%%%%%%%%%%%%%%%%%%%%%%%%%%%%%%%%%%%%%%%%%%%%%%%%%%%%%%%%%%%%%%%%%%%

\begin{table}[p]
\begin{center}
\begin{tabular}{||c|c|c|c|c|c||c||c||}
\hline
$x_{{\rm Bj}}$ bin & $d\sigma/dx_{{\rm Bj}}$ & $\delta_{stat}$ 
& $\delta_{syst}$ & $\delta_{CAL}$  & $\delta_{FPC}$&\footnotesize $C_{QED}$&\footnotesize $C_{HAD}$\normalsize\\
&$\rm{(nb)}$ & &  &  & & & \normalsize\\

\hline\hline
0.0004 - 0.001&  $275$  & $\pm 6$ & $^{+19}_{-19}$ & $^{+16}_{-20}$ & $^{+34}_{-35}$& $0.97$ & $0.86$ \\
0.001 - 0.002&   $200$  & $\pm 4$ & $^{+10}_{-9} $  & $^{+14}_{-14}$ & $^{+23}_{-20}$ & $0.97$ &$0.88$ \\
0.002 - 0.003&   $125$  & $\pm 3$ & $^{+2}_{-3}  $   & $^{+6}_{-9}$ & $^{+11}_{-12}$ & $0.96$ & $0.90$ \\
0.003 - 0.004&   $89$ & $\pm 2$ & $^{+4}_{-3}    $   & $^{+5}_{-6}$ & $^{+9}_{-9}$ & $0.95$ & $0.92$ \\
0.004 - 0.005&   $65$ & $\pm 2$ & $^{+1}_{-1}     $  & $^{+4}_{-5}$ & $^{+6}_{-6}$ & $0.94$ & $0.95$ \\
\hline
\end{tabular}
\caption{ The differential cross section, $d\sigma/dx_{{\rm Bj}}$, in bins of 
$x_{{\rm Bj}}$ for inclusive forward jets. The statistical, systematic and 
jet-energy-scale uncertainties are shown separately (see the caption of Table~\ref{tab-1}).}
\label{tab-2}
\end{center}
\end{table}
%%%%%%%%%%%%%%%%%%%%%%%%%%%%%%%%%%%%%%%%%%%%%%%%%%%%%%%%%%%%%%%%%%%%%%%%%%%%%%%%

\begin{table}[p]
\begin{center}
\begin{tabular}{||c|c|c|c|c|c||c||c||}
\hline
$E_{\rm T}^{\rm jet}$ bin & $d\sigma/dE_{\rm T}^{\rm jet}$ & $\delta_{stat}$ 
& $\delta_{syst}$ & $\delta_{CAL}$ & $\delta_{FPC}$&\footnotesize $C_{QED}$&\footnotesize $C_{HAD}$\normalsize\\
${\rm (GeV)}$ &$\rm{(pb}/{\rm GeV)}$&  &  &  & & & \normalsize\\

\hline\hline
5 - 6.5&   $245.1$ & $\pm 3.2$  & $^{+6.9}_{-7.6}$     & $^{+15.0}_{-18.7}$ & $^{+32.6}_{-31.3}$ & $0.96$ &$0.91$ \\
6.5 - 8&   $123.2$ & $\pm 2.4$  & $^{+4.5}_{-4.5}$     & $^{+7.7}_{-8.1}$& $^{+10.0}_{-10.2}$ & $0.97$ & $0.92$ \\
8 - 9.5&   $42.04$ & $\pm 1.40$ & $^{+1.22}_{-1.20}$ & $^{+1.9}_{-2.35}$ & $^{+1.50}_{-1.69}$& $0.96$ & $0.89$ \\
9.5 - 11&  $13.38$ & $\pm 0.73$ & $^{+1.17}_{-0.99}$ & $^{+0.63}_{-1.49}$& $^{+0.62}_{-1.35}$ & $0.96$ & $0.87$ \\
11 - 14&   $ 2.21$ & $\pm 0.21$ & $^{+0.16}_{-0.20}$ & $^{+0.05}_{-0}$ & $^{+0.08}_{-0.21}$& $0.93$ & $0.89$ \\
\hline
\end{tabular}
\caption{ The differential cross section, $d\sigma/dE_{T}^{\rm jet}$, in bins of 
$E_{T}^{\rm jet}$ for inclusive forward jets. The statistical, systematic and 
jet-energy-scale uncertainties are shown separately (see the caption of Table~\ref{tab-1}).}
\label{tab-2}
\end{center}
\end{table}
%%%%%%%%%%%%%%%%%%%%%%%%%%%%%%%%%%%%%%%%%%%%%%%%%%%%%%%%%%%%%%%%%%%%%%%%%%%%%%%%

\begin{table}[p]
\begin{center}
\begin{tabular}{||c|c|c|c|c|c||c||c||}
\hline
${\eta}^{\rm jet}$ bin & $d\sigma/d{\eta}^{\rm jet}$ & $\delta_{stat}$ 
& $\delta_{syst}$ & $\delta_{CAL}$  & $\delta_{FPC}$&\footnotesize $C_{QED}$&\footnotesize $C_{HAD}$\normalsize\\
&$\rm{(pb)}$ & & &  & & & \normalsize\\
\hline\hline
2 - 2.3&    $118.0$ & $\pm 5.1$ & $^{+7.8}_{-7.8}$       & $^{+8.4}_{-6.8}$     & $^{+0}_{-0.1}$   & $0.97$ & $0.87$ \\
2.3 - 2.6&  $361.4 $ & $\pm 8.5 $ & $^{+20.0}_{-21.0}$ & $^{+24.2}_{-34.1}$  & $^{+0}_{-1.3}$& $0.95$ & $0.90$ \\
2.6 - 2.9&  $438.2$ & $\pm 9.6$ & $^{+14.9}_{-12.4}$   & $^{+40.9}_{-48.7}$  & $^{+0.8}_{-1.0}$& $0.96$ & $0.93$ \\
2.9 - 3.5&  $331.0$ & $\pm 8.1$ & $^{+19.2}_{-20.2}$   & $^{+17.0}_{-25.2}$  & $^{+15.5}_{-15.4}$& $0.95$ & $0.92$ \\
3.5 - 4.3&  $211.6$ & $\pm 4.2$ & $^{+6.6}_{-7.7}$       & $^{+4.8}_{-5.1}$ & $^{+54.9}_{-79.1}$ & $0.98$ & $1.01$ \\
\hline
\end{tabular}
\caption{ The differential cross section, $d\sigma/d{\eta}^{\rm jet}$, in bins of 
$\eta^{\rm jet}$ for inclusive forward jets. The statistical, systematic and 
jet-energy-scale uncertainties are shown separately (see the caption of Table~\ref{tab-1}).}
\label{tab-4}
\end{center}
\end{table}
%%%%%%%%%%%%%%%%%%%%%%%%%%%%%%%%%%%%%%%%%%%%%%%%%%%%%%%%%%%%%%%%%%%%%%%%%%%%%%%%

\begin{table}[p]
\begin{center}
\begin{tabular}{||c|c|c|c|c|c||c||c||}
\hline
$\Delta\eta_1$ bin & $d\sigma/d\Delta{\eta}_1$& $\delta_{stat}$ 
& $\delta_{syst}$ & $\delta_{CAL}$  & $\delta_{FPC}$&\footnotesize $C_{QED}$&\footnotesize $C_{HAD}$\normalsize\\
&$\rm{(pb)}$ & & &  &  & & \normalsize\\
\hline\hline
0.0 - 0.7&  $82.88$ & $\pm 3.77$ & $^{+12.88}_{-12.88}$ & $^{+8.24}_{-10.9}$ & $^{+4.24}_{-4.24}$ & $0.98$ & $0.74$ \\
0.7 - 1.4&  $79.35$ & $\pm 3.65$ & $^{+11.18}_{-11.01}$ & $^{+9.40}_{-8.66}$  & $^{+5.18}_{-4.13}$& $0.99$ & $0.76 $ \\
1.4 - 2.1&  $50.68$ & $\pm 2.62$ & $^{+5.53}_{-6.12}$    & $^{+4.79}_{-6.91} $ & $^{+3.68}_{-3.49}$ & $0.98$ & $0.76$ \\
2.2 - 4  &  $15.30$  & $\pm 0.99$  & $^{+2.35}_{-1.67}$   & $^{+1.16}_{-2.42}$  & $^{+1.36}_{-2.10}$& $0.96$ & $0.75$ \\
\hline
\end{tabular}
\caption{ The differential cross section, $d\sigma/d\Delta{\eta}_1$, in bins of 
$\Delta{\eta}_1$ for ``forward jet+dijet'' events. The statistical, systematic and 
jet-energy-scale uncertainties are shown separately (see the caption of Table~\ref{tab-1}).}
\label{tab-5}
\end{center}
\end{table}
%%%%%%%%%%%%%%%%%%%%%%%%%%%%%%%%%%%%%%%%%%%%%%%%%%%%%%%%%%%%%%%%%%%%%%%%%%%%%%%%

\begin{table}[p]
\begin{center}
\begin{tabular}{||c|c|c|c|c|c||c||c||}
\hline
$\Delta\eta_2$ bin & $d\sigma/d\Delta{\eta}_1$& $\delta_{stat}$ 
& $\delta_{syst}$ & $\delta_{CAL}$ & $\delta_{FPC}$ &\footnotesize $C_{QED}$&\footnotesize $C_{HAD}$\normalsize\\
&$\rm{(pb)}$ & & &  &  & & \normalsize\\
\hline\hline
0.0 - 0.8&  $56.38$ & $\pm 2.64$ & $^{+5.26}_{-4.72}$ & $^{+5.42}_{-5.97}$ & $^{+1.35}_{-1.42}$& $0.98$ & $0.78$ \\
0.8 - 1.6&  $76.23$ & $\pm 3.33$ & $^{+5.23}_{-6.18}$ & $^{+7.44}_{-9.52}$ & $^{+2.51}_{-3.09}$& $0.96$ & $0.79$ \\
1.6 - 2.4&  $56.06$ & $\pm 2.82$ & $^{+9.33}_{-8.69}$ & $^{+5.18}_{-7.31}$ & $^{+4.44}_{-4.79}$& $1.01$ & $0.75$ \\
2.4 - 3.2&  $19.44$ & $\pm 1.27$ & $^{+4.46}_{-3.98}$ & $^{+2.12}_{-2.70}$ & $^{+2.69}_{-2.58}$& $0.97$ & $0.69$ \\
\hline
\end{tabular}
\caption{ The differential cross section, $d\sigma/d\Delta{\eta}_2$, in bins of 
$\Delta{\eta}_2$ for ``forward jet+dijet'' events. The statistical, systematic and 
jet-energy-scale uncertainties are shown separately (see the caption of Table~\ref{tab-1}).}
\label{tab-6}
\end{center}
\end{table}
%%%%%%%%%%%%%%%%%%%%%%%%%%%%%%%%%%%%%%%%%%%%%%%%%%%%%%%%%%%%%%%%%%%%%%%%%%%%%%%%

\begin{table}[p]
\begin{center}
\begin{tabular}{||c|c|c|c|c|c||c||c||}
\hline
$\Delta\eta_2$ bin & $d\sigma/d\Delta{\eta}_1$& $\delta_{stat}$ 
& $\delta_{syst}$ & $\delta_{CAL}$  & $\delta_{FPC}$ &\footnotesize $C_{QED}$&\footnotesize $C_{HAD}$\normalsize\\
&$\rm{(pb)}$ & & &  & & & \normalsize\\
\hline\hline
0.0 - 0.8&  $14.86$ & $\pm 1.36$  & $^{+1.17}_{-1.71} $ & $^{+1.65}_{-1.86}$  & $^{+0}_{-0.24}$ & $1.00$ & $0.82$ \\
0.8 - 1.6&  $30.04$ & $\pm 2.17$  & $^{+4.58}_{-5.04} $ & $^{+3.21}_{-3.04}$ & $^{+0.26}_{-0.34}$  & $0.99$ & $0.81$ \\
1.6 - 2.4&  $29.48$ & $\pm 1.93$  & $^{+4.13}_{-3.78} $ & $^{+2.64}_{-3.42}$  & $^{+1.12}_{-0.79}$ & $1.02$ & $0.78$ \\
2.4 - 3.4&  $12.33$  & $\pm 0.94$ & $^{+2.59}_{-2.27} $ & $^{+1.44}_{-1.78}$  & $^{+1.63}_{-1.50}$ & $0.96$ & $0.68$ \\
\hline
\end{tabular}
\caption{ The differential cross section, $d\sigma/d\Delta{\eta}_2$, in bins of 
$\Delta{\eta}_2$ for ``forward jet+dijet'' events in the case of $\Delta\eta_1 < 1$. 
The statistical, systematic and jet-energy-scale uncertainties are shown separately (see the caption of Table~\ref{tab-1}).}
\label{tab-7}
\end{center}
\end{table}
%%%%%%%%%%%%%%%%%%%%%%%%%%%%%%%%%%%%%%%%%%%%%%%%%%%%%%%%%%%%%%%%%%%%%%%%%%%%%%%%

\begin{table}[p]
\begin{center}
\begin{tabular}{||c|c|c|c|c|c||c||c||}
\hline
$\Delta{\eta}_2$ bin & $d\sigma/d\Delta{\eta}_1$& $\delta_{stat}$ 
& $\delta_{syst}$ & $\delta_{CAL}$  & $\delta_{FPC}$&\footnotesize $C_{QED}$&\footnotesize $C_{HAD}$\normalsize\\
&$\rm{(pb)}$ & & & & & & \normalsize\\
\hline\hline
0.0 - 0.6&  $40.03$ & $\pm 2.28$ & $^{+5.32}_{-4.81}$ & $^{+3.62}_{-4.12}$ & $^{+1.19}_{-1.14}$ & $0.97$ & $0.75$ \\
0.6 - 1.2&  $46.58$ & $\pm 2.59$ & $^{+1.66}_{-2.71}$ & $^{+3.87}_{-5.97}$ & $^{+1.58}_{-1.76}$ & $0.96$ & $0.80$ \\
1.2 - 1.8&  $36.46$ & $\pm 2.74$ & $^{+7.01}_{-5.80}$ & $^{+4.08}_{-5.49}$ & $^{+4.38}_{-5.49}$ & $0.98$ & $0.76$ \\
1.8 - 2.8&  $10.27$ & $\pm 1.03$ & $^{+2.91}_{-2.87}$ & $^{+0.80}_{-1.44}$ & $^{+1.91}_{-2.51}$ & $0.99$ & $0.71$ \\
\hline
\end{tabular}
\caption{ The differential cross section, $d\sigma/d\Delta{\eta}_2$, in bins of 
$\Delta{\eta}_2$ for ``forward jet+dijet'' events in the case of $\Delta\eta_1 > 1$. 
The statistical, systematic and jet-energy-scale uncertainties are shown separately (see the caption of Table~\ref{tab-1}).}
\label{tab-8}
\end{center}
\end{table}
%%%%%%%%%%%%%%%%%%%%%%%%%%%%%%%%%%%%%%%%%%%%%%%%%%%%%%%%%%%%%%%%%%%%%%%%%%%%%%%%

\begin{table}[p]
\begin{center}
\begin{tabular}{||c|c|c|c|c||c||c||}
\hline
$(E_{\rm T}^{\rm jet})^2$ & ${\eta}^{\rm jet}$ bin & $d^3\sigma/dQ^2d(E_T^{jet})^2d{\eta}^{\rm
jet}\pm\delta_{stat}\pm\delta_{syst}$  & $\delta_{CAL}$  & $\delta_{FPC}$ &\footnotesize $C_{QED}$&\footnotesize $C_{HAD}$\normalsize\\
$\rm{(GeV^2)}$& &$\rm{(nb/GeV}^4)$ &  &  && \normalsize\\
\hline\hline
\multicolumn{7}{||c||} {$20 < Q^2 < 40 ~{\rm ( GeV}^2$)} \\
\hline
        &2.4-2.7&  $525\pm 20^{+61}_{-62}$    & $^{+42}_{-63}$ & - & $0.93$ & $0.88$ \\
25-36   &2.7-3.1&  $658\pm 24^{+50}_{-44}$    & $^{+52}_{-90}$ & $^{+1}_{-2}$& $0.94$ & $0.91$ \\
        &3.1-3.7&  $463\pm 20^{+20}_{-31}$    & $^{+22}_{-10}$ & $^{+66}_{-143}$& $0.98$ & $0.86$ \\
        &3.7-4.3&  $359\pm 13^{+13}_{-18}$    & $^{+7}_{-11}$  & $^{+111}_{-197}$& $0.96$ & $0.93$ \\

        &2.0-2.4&  $133\pm 4^{+10}_{-10}$     &$^{+10}_{-12}$ & - & $0.96$ & $0.95$ \\
        &2.4-2.7&  $200\pm 6^{+13}_{-13}$     &$^{+15}_{-18}$ & - & $0.98$ & $0.89$ \\
36-100 &2.7-3.1&   $170\pm 6^{+6}_{-6}$       &$^{+17}_{-14}$ &$^{+2}_{-1}$ & $0.97$ & $0.87$ \\
        &3.1-3.7&  $106\pm 4^{+6}_{-5}$       &$^{+6}_{-7}$ &$^{+12}_{-15}$& $0.97$ & $0.89$ \\
        &3.7-4.3&  $73.2\pm 2.6^{+2.6}_{-2.7}$      &$^{+1.7}_{-1.2}$ &$^{+20.9}_{-27.7}$ & $1.00$ & $0.88$ \\

        &2.0-2.4&  $17.8\pm 0.8^{+0.7}_{-0.7}$   &$^{+1.2}_{-1.4}$ & - & $0.99$ & $0.95$ \\
        &2.4-2.7&  $13.3\pm 0.8^{+0.8}_{-0.7}$   &$^{+1.4}_{-1.5}$ & - & $0.99$ & $0.94$ \\
100-400 &2.7-3.1&  $9.56\pm 0.68^{+0.42}_{-0.27}$ &$^{+0.82}_{-0.95}$ & - & $0.96$ & $0.92$ \\
        &3.1-3.7&  $6.14\pm 0.58^{+0.95}_{-0.64}$     &$^{+0.41}_{-0.87}$ &$^{+0.86}_{-1.55}$ & $0.97$ & $0.98$ \\
        &3.7-4.3&  $2.66\pm 0.25^{+0.18}_{-0.16}$     & - &$^{+0.83}_{-1.48}$& $0.98$ & $0.93$ \\
\hline\hline
\multicolumn{7}{||c||} {$40 < Q^2 < 100 ~{\rm ( GeV}^2$)}\\
\hline

        &2.4-2.7&  $83.6\pm 5.0^{+3.1}_{-3.2}$     &$^{+4.6}_{-7.1}$ &$^{+0}_{-0.2}$& $0.89$ & $0.90$ \\
25-36   &2.7-3.1&  $105\pm 5^{+6}_{-4}$&$^{+8}_{-11}$ & - & $0.94 $ & $0.90$ \\
        &3.1-3.7&  $78.3\pm 4.7^{+4.2}_{-4.2}$     &$^{+2.0}_{-4.5}$ &$^{+7.8}_{-12.0}$& $0.95$ & $0.89$ \\
        &3.7-4.3&  $57.2\pm 3.0^{+6.1}_{-5.6}$     &$^{+0.4}_{-0.9}$ &$^{+15.7}_{-26.3}$ & $0.94$ & $0.85$ \\

        &2.0-2.4&  $24.2\pm 10.0^{+1.9}_{-1.9}$   &$^{+1.5}_{-1.4}$ & - & $0.94$ & $0.92$ \\
        &2.4-2.7&  $35.8\pm 1.4^{+2.1}_{-2.2}$   &$^{+3.1}_{-2.3}$ & - & $0.98$ & $0.93$ \\
36-100   &2.7-3.1& $26.5\pm 1.2^{+1.5}_{-1.3}$&$^{+2.2}_{-2.2}$ &$^{+0.1}_{-0.2}$ & $0.96$ & $0.88$ \\
        &3.1-3.7&  $19.8\pm 0.9^{+1.4}_{-1.5}$   &$^{+0.9}_{-0.9}$ &$^{+1.4}_{-1.5}$& $0.96$ & $0.89$ \\
        &3.7-4.3&  $13.7\pm 0.6^{+0.4}_{-0.4}$   &$^{+0.2}_{-0.2}$ &$^{+3.1}_{-4.1}$& $0.99$ & $0.94$ \\

        &2.0-2.4&  $3.59\pm 0.18^{+0.21}_{-0.22}$      &$^{+0.29}_{-0.30}$ & - & $0.97$ & $0.97$ \\
        &2.4-2.7&  $2.57\pm 0.17^{+0.10}_{-0.06}$      &$^{+0.17}_{-0.33}$ & - & $0.96$ & $0.97$ \\
100-400 &2.7-3.1&  $1.71\pm 0.13^{+0.11}_{-0.19}$      &$^{+0.15}_{-0.15}$ & - & $0.96$ & $0.94$ \\
        &3.1-3.7&  $0.99\pm 0.08^{+0.16}_{-0.18}$      & - &$^{+0.13}_{-0.15}$& $0.96$ & $0.96$ \\
        &3.7-4.3&  $0.58\pm 0.05^{+0.05}_{-0.06}$      & - &$^{+0.19}_{-0.30}$ & $0.97$ & $0.98$ \\
\hline
\end{tabular}
\caption{ The differential cross sections as a function of
$\eta^{jet}$ in different bins of $Q^2$ and $(E_{T}^{jet})^2$ for inclusive forward jets.
The jet-energy-scale uncertainties are shown separately (see the caption of Table~\ref{tab-1}).}
\label{tab-4}
\end{center}
\end{table}

\newpage

%s
\newpage
\newpage
\newpage
\newpage
\newpage
\newpage
\newpage
\newpage
\newpage

\end{document}

%% file: zeus_def.tex
\newcommand{\ZcoosysB}{%
The ZEUS coordinate system is a right-handed Cartesian system, with the $Z$
axis pointing in the proton beam direction, referred to as the ``forward
direction'', and the $X$ axis pointing left towards the centre of HERA.
The coordinate origin is at the nominal interaction point.\xspace}
\newcommand{\Zpsrap}{}

\newcommand{\ZcoosysfnBeta}{\footnote{\ZcoosysB\Zpsrap}}
\newcommand{\Zdetdesc}{%
A detailed description of the ZEUS detector can be found 
elsewhere~\cite{zeus:1993:bluebook}. A brief outline of the 
components that are most relevant for this analysis is given
below.\xspace}
\newcommand{\Zctddesc}[1]{%
Charged particles are tracked in the central tracking detector (CTD)~\citeCTD,
which operates in a magnetic field of $1.43\Tesla$ provided by a thin 
superconducting solenoid. The CTD consists of 72~cylindrical drift chamber 
layers, organised in nine superlayers covering the polar-angle region 
\mbox{$15^\circ<\theta<164^\circ$}. The transverse-momentum resolution for
full-length tracks is $\sigma(p_T)/p_T=0.0058p_T\oplus0.0065\oplus0.0014/p_T$,
with $p_T$ in $\Gev$.}
\newcommand{\Zcaldesc}{%
The high-resolution uranium--scintillator calorimeter (CAL)~\citeCAL consists 
of three parts: the forward (FCAL), the barrel (BCAL) and the rear (RCAL)
calorimeters. Each part is subdivided transversely into towers and
longitudinally into one electromagnetic section (EMC) and either one (in RCAL)
or two (in BCAL and FCAL) hadronic sections (HAC). The smallest subdivision of
the calorimeter is called a cell.  The CAL energy resolutions, as measured under
test-beam conditions, are $\sigma(E)/E=0.18/\sqrt{E}$ for electrons and
$\sigma(E)/E=0.35/\sqrt{E}$ for hadrons, with $E$ in $\Gev$.
}
\chardef\usc=95
\chardef\til=126
\catcode`\@=11 % @ signs are now treated as letters
\DeclareRobustCommand\xdotspace{\futurelet\@let@token\@xdotspace}
\def\@xdotspace{%
  \ifx\@let@token.\else
  \ifx\@let@token\bgroup.\else
  \ifx\@let@token\egroup.\else
  \ifx\@let@token\/.\else
  \ifx\@let@token\ .\else
  \ifx\@let@token~.\else
  \ifx\@let@token!.\else
  \ifx\@let@token,.\else
  \ifx\@let@token:.\else
  \ifx\@let@token;.\else
  \ifx\@let@token?.\else
  \ifx\@let@token/.\else
  \ifx\@let@token'.\else
  \ifx\@let@token).\else
  \ifx\@let@token-.\else
  \ifx\@let@token\@xobeysp.\else
  \ifx\@let@token\space.\else
  \ifx\@let@token\@sptoken.\else
   .\space
   \fi\fi\fi\fi\fi\fi\fi\fi\fi\fi\fi\fi\fi\fi\fi\fi\fi\fi}
\catcode`\@=12 % @ signs are no longer letters

\newcommand{\stru}[2]{%
   \relax\ifmmode\hbox{\vrule height#1 depth#2 width0pt}%
   \else\vrule height#1 depth#2 width0pt\fi}
\newcommand{\Ronum}[1]{\uppercase\expandafter{\romannumeral#1}}
\newcommand{\ronum}[1]{\expandafter{\romannumeral#1}}
\DeclareRobustCommand{\LaTeXZ}{%
  \LaTeX\kern-.05em4\kern-.1em
  {\raisebox{-0.2ex}{$\scriptstyle\text{ZEUS}$}}\xspace}
\DeclareMathAlphabet{\mathbf}{OT1}{cmr}{bx}{sl}
\newcommand{\eVdist}{\kern-0.06667em}

\newcommand{\Gev}{{\text{Ge}\eVdist\text{V\/}}}

\newcommand{\gev}{{\,\text{Ge}\eVdist\text{V\/}}}

\newcommand{\Tesla}{\,\text{T}}

\newcommand{\slashfrac}[2]{%
  \raisebox{0.5ex}{\ensuremath #1}\kern-0.12em/\kern-0.08em
  \raisebox{-.8ex}{\ensuremath #2}}
\newcommand{\sqr}[3]{%
    {\vcenter{\hrule height.#3ex\hbox{\vrule width.#2ex height#1ex
     \kern#1ex\vrule width.#3ex}\hrule height.#2ex}}}

\catcode`\@=11 % @ signs are now treated as letters
\newcommand{\parenbar}{\mathpalette\p@renb@r}
\def\p@renb@r#1#2{\vbox{%
  \ifx#1\scriptscriptstyle \dimen@.7em\dimen@ii.2em\else
  \ifx#1\scriptstyle \dimen@.8em\dimen@ii.25em\else
  \dimen@1em\dimen@ii.4em\fi\fi \offinterlineskip
  \ialign{\hfill##\hfill\cr
    \vbox{\hrule width\dimen@ii}\cr
    \noalign{\vskip-.3ex}%
    \hbox to\dimen@{$\mathchar300\hfil\mathchar301$}\cr
    \noalign{\vskip-.3ex}%
    $#1#2$\cr}}}
\catcode`\@=12 % @ signs are no longer letters

\newcommand{\IP}{{\rm I$\kern-0.01667em$P}\xspace}

\newcommand{\data}{{\rm data}}

\mathchardef\qsm=63
\mathchardef\pls=43
\mathchardef\mns=512
\mathchardef\plm=518
\mathchardef\eql=61
\mathchardef\smallleft=300
\mathchardef\smallright=301
\mathchardef\les=316
\mathchardef\gre=318
\mathchardef\leq=532
\mathchardef\grq=533
\catcode`\@=11 % @ signs are now treated as letters
\newcounter{pict@width}
\newcounter{pict@height}
\newlength{\pict@scale}
\newcommand{\psfigadd}[4]{%
\setcounter{pict@width}{1*\ratio{#2+\pict@scale/2}{\pict@scale}}
\setcounter{pict@height}{1*\ratio{#3+\pict@scale/2}{\pict@scale}}
\setlength{\unitlength}{\pict@scale}
\hbox to #2{\hspace{-\fill}\begin{picture}(\thepict@width,\thepict@height)
\put(0,0){\psfig{figure=#1,width=#2,height=#3,clip=}}
\SetScale{0.283466457}
\SetWidth{1.763889}
{#4}
\end{picture}}
}
\newcounter{pict@widthfst}
\newcounter{pict@widthscd}
\newcounter{pict@widthtot}
\newcommand{\psfigaddtwo}[7]{%
\setcounter{pict@widthfst}{1*\ratio{#2+\pict@scale/2}{\pict@scale}}
\setcounter{pict@widthscd}{1*\ratio{#2+#4+\pict@scale/2}{\pict@scale}}
\setcounter{pict@widthtot}{1*\ratio{#2+#4+#6+\pict@scale/2}{\pict@scale}}
\setcounter{pict@height}{1*\ratio{#3+\pict@scale/2}{\pict@scale}}
\setlength{\unitlength}{\pict@scale}
\hbox{\hspace{-\fill}\begin{picture}(\thepict@widthtot,\thepict@height)
\put(0,0){\psfig{figure=#1,width=#2,height=#3,clip=}}
\put(\thepict@widthscd,0){\psfig{figure=#5,width=#6,height=#3,clip=}}
\SetScale{0.283466457}
\SetWidth{1.763889}
{#7}
\end{picture}}
}
\newcommand{\psfigror}[4]{%
\setcounter{pict@width}{1*\ratio{#2+\pict@scale/2}{\pict@scale}}
\setcounter{pict@height}{1*\ratio{#3+\pict@scale/2}{\pict@scale}}
\setlength{\unitlength}{\pict@scale}
\hbox{\begin{picture}(\thepict@width,\thepict@height)
\put(0,\thepict@height){\psfig{figure=#1,width=#3,height=#2,clip=,angle=270}}
\SetScale{0.283466457}
\SetWidth{1.763889}
{#4}
\end{picture}}
}
\newcommand{\psfigrol}[4]{%
\setcounter{pict@width}{1*\ratio{#2+\pict@scale/2}{\pict@scale}}
\setcounter{pict@height}{1*\ratio{#3+\pict@scale/2}{\pict@scale}}
\setlength{\unitlength}{\pict@scale}
\hbox{\begin{picture}(\thepict@width,\thepict@height)
\put(0,0){\psfig{figure=#1,width=#3,height=#2,clip=,angle=90}}
\SetScale{0.283466457}
\SetWidth{1.763889}
{#4}
\end{picture}}
}
\catcode`\@=12 % @ signs are no longer letters
\newlength\listtextwidth

\catcode`\@=11 % @ signs are now treated as letters
\newlength{\@tabfninsert}
\newlength{\@tabfnwidth}
\newcommand{\tabfootnote}[2]{%
  \setlength{\@tabfninsert}{0.8em}
  \setlength{\@tabfnwidth}{\textwidth}
  \addtolength{\@tabfnwidth}{-\@tabfninsert}
  \addtolength{\@tabfnwidth}{-0.4em}
  \noindent\makebox[\@tabfninsert][r]{\footnotesize$^{#1}$\hfil}\hfill%
  \parbox[t]{\@tabfnwidth}{\footnotesize #2\hfill}}
\catcode`\@=12 % @ signs are no longer letters

%% file: DESY-07-100.bbl
\begin{mcbibliography}{10}

\bibitem{sovjnp:15:438}
V.N.~Gribov and L.N.~Lipatov,
\newblock Sov.\ J.\ Nucl.\ Phys.{} {\bf 15},~438~(1972)\relax
\relax
\bibitem{jetp:46:641}
Yu.L.~Dokshitzer,
\newblock Sov.\ Phys.\ JETP{} {\bf 46},~641~(1977)\relax
\relax
\bibitem{np:b126:298}
G.~Altarelli and G.~Parisi,
\newblock Nucl.\ Phys.{} {\bf B~126},~298~(1977)\relax
\relax
\bibitem{sovjnp:28:822}
Ya.Ya.~Balitski\u i and L.N.~Lipatov,
\newblock Sov.\ J.\ Nucl.\ Phys.{} {\bf 28},~822~(1978)\relax
\relax
\bibitem{np:b296:49}
M.~Ciafaloni,
\newblock Nucl.\ Phys.{} {\bf B~296},~49~(1988)\relax
\relax
\bibitem{np:b336:18}
S.~Catani, F.~Fiorani and G.~Marchesini,
\newblock Nucl.\ Phys.{} {\bf B~336},~18~(1990)\relax
\relax
\bibitem{npps:18c:125}
A.H.~Mueller,
\newblock Nucl.\ Phys.\ Proc.\ Suppl.{} {\bf 18~C},~125~(1991)\relax
\relax
\bibitem{epj:c6:239}
ZEUS \coll, J.~Breitweg \etal,
\newblock Eur.\ Phys.\ J.{} {\bf C~6},~239~(1999)\relax
\relax
\bibitem{pl:b474:223}
ZEUS \coll, J.~Breitweg \etal,
\newblock Phys.\ Lett.{} {\bf B~474},~223~(2000)\relax
\relax
\bibitem{np:b538:3}
H1 \coll, C.~Adloff \etal,
\newblock Nucl.\ Phys.{} {\bf B~538},~3~(1999)\relax
\relax
\bibitem{cpc:86:147}
H.~Jung,
\newblock Comp.\ Phys.\ Comm.{} {\bf 86},~147~(1995)\relax
\relax
\bibitem{pl:b175:453}
G.~Gustafson,
\newblock Phys.\ Lett.{} {\bf B~175},~453~(1986)\relax
\relax
\bibitem{np:b306:746}
G.~Gustafson and U.~Petterson,
\newblock Nucl.\ Phys.{} {\bf B~306},~746~(1988)\relax
\relax
\bibitem{zfp:c43:625}
B.~Andersson et al.,
\newblock Z.\ Phys.{} {\bf C~43},~625~(1989)\relax
\relax
\bibitem{np:b485:291}
S.~Catani and M.H.~Seymour,
\newblock Nucl.\ Phys.{} {\bf B~485},~579~(1998)\relax
\relax
\bibitem{nlojet}
Z.~Nagy and Z.~Trocsanyi,
\newblock Phys. Rev. Lett.{} {\bf 87},~082001~(2001)\relax
\relax
\bibitem{desy-05-117}
ZEUS \coll, S.~Chekanov \etal,
\newblock Phys. Lett.{} {\bf B~632},~13~(2006)\relax
\relax
\bibitem{desy-05-135}
H1 \coll, A.~Aktas \etal,
\newblock Eur. Phys. J.{} {\bf C 46},~27~(2006)\relax
\relax
\bibitem{epj:c19:351}
H.~Jung and G.~P.~Salam,
\newblock Eur.\ Phys.\ J.{} {\bf C~19}~(2001)\relax
\relax
\bibitem{zeus:1993:bluebook}
ZEUS \coll, U.~Holm~(ed.),
\newblock {\em The {ZEUS} Detector}.
\newblock Status Report (unpublished), DESY (1993),
\newblock available on
  \texttt{http://www-zeus.desy.de/bluebook/bluebook.html}\relax
\relax
\bibitem{nim:a279:290}
N.~Harnew \etal,
\newblock Nucl.\ Inst.\ Meth.{} {\bf A~279},~290~(1989)\relax
\relax
\bibitem{npps:b32:181}
B.~Foster \etal,
\newblock Nucl.\ Phys.\ Proc.\ Suppl.{} {\bf B~32},~181~(1993)\relax
\relax
\bibitem{nim:a338:254}
B.~Foster \etal,
\newblock Nucl.\ Inst.\ Meth.{} {\bf A~338},~254~(1994)\relax
\relax
\bibitem{nim:a309:77}
M.~Derrick \etal,
\newblock Nucl.\ Inst.\ Meth.{} {\bf A~309},~77~(1991)\relax
\relax
\bibitem{nim:a309:101}
A.~Andresen \etal,
\newblock Nucl.\ Inst.\ Meth.{} {\bf A~309},~101~(1991)\relax
\relax
\bibitem{nim:a321:356}
A.~Caldwell \etal,
\newblock Nucl.\ Inst.\ Meth.{} {\bf A~321},~356~(1992)\relax
\relax
\bibitem{nim:a336:23}
A.~Bernstein \etal,
\newblock Nucl.\ Inst.\ Meth.{} {\bf A~336},~23~(1993)\relax
\relax
\bibitem{desy-92-066}
J.~Andruszk\'ow \etal,
\newblock Preprint \mbox{DESY-92-066}, DESY, 1992\relax
\relax
\bibitem{zfp:c63:391}
ZEUS \coll, M.~Derrick \etal,
\newblock Z.\ Phys.{} {\bf C~63},~391~(1994)\relax
\relax
\bibitem{acpp:b32:2025}
J.~Andruszk\'ow \etal,
\newblock Acta Phys.\ Pol.{} {\bf B~32},~2025~(2001)\relax
\relax
\bibitem{fpcnim:a450:235}
A.~Bamberger \etal,
\newblock Nucl.\ Inst.\ Meth.{} {\bf A~450},~235~(2000)\relax
\relax
\bibitem{hep-ex-0208037}
ZEUS \coll, S.~Chekanov \etal,
\newblock Phys. Lett.{} {\bf B547},~164~(2002)\relax
\relax
\bibitem{nim:a365:508}
H.~Abramowicz, A.~Caldwell and R.~Sinkus,
\newblock Nucl.\ Inst.\ Meth.{} {\bf A~365},~508~(1995)\relax
\relax
\bibitem{nim:a391:360}
R.~Sinkus and T.~Voss,
\newblock Nucl.\ Inst.\ Meth.{} {\bf A~391},~360~(1997)\relax
\relax
\bibitem{proc:hera:1991:23}
S.~Bentvelsen, J.~Engelen and P.~Kooijman,
\newblock {\em Proc.\ Workshop on Physics at {HERA}}, W.~Buchm\"uller and
  G.~Ingelman~(eds.), Vol.~1, p.~23.
\newblock Hamburg, Germany, DESY (1992)\relax
\relax
\bibitem{proc:hera:1991:43}
K.C.~H\"oger,
\newblock {\em Proc.\ Workshop on Physics at {HERA}}, W.~Buchm\"uller and
  G.~Ingelman~(eds.), Vol.~1, p.~43.
\newblock Hamburg, Germany, DESY (1992)\relax
\relax
\bibitem{epj:c1:81}
ZEUS \coll, J.~Breitweg \etal,
\newblock Eur.\ Phys.\ J.{} {\bf C~1},~81~(1998)\relax
\relax
\bibitem{np:b406:187}
S.~Catani \etal,
\newblock Nucl.\ Phys.{} {\bf B~406},~187~(1993)\relax
\relax
\bibitem{pr:d48:3160}
S.D.~Ellis and D.E.~Soper,
\newblock Phys.\ Rev.{} {\bf D~48},~3160~(1993)\relax
\relax
\bibitem{zpf:c2:237}
K.H.~Streng, T.F.~Walsh, P.M.~Zerwas,
\newblock Z.\ Phys.{} {\bf C~2},~237~(1979)\relax
\relax
\bibitem{cpc:101:108}
G.~Ingelman, A.~Edin and J.~Rathsman,
\newblock Comp.\ Phys.\ Comm.{} {\bf 101},~108~(1997)\relax
\relax
\bibitem{cpc:71:15}
L.~L\"onnblad,
\newblock Comp.\ Phys.\ Comm.{} {\bf 71},~15~(1992)\relax
\relax
\bibitem{cpc:143:100}
H.~Jung,
\newblock Comp.\ Phys.\ Comm.{} {\bf 143},~100~(2002)\relax
\relax
\bibitem{spb:2004ugd}
H.~Jung,
\newblock {\em Strbske Pleso 2004, Deep inelastic scattering}, D.~Bruncko et
  al.~(ed.), Vol.~1, pp.~299--302.
\newblock Institute of Experimental Physics SAS, Kosi\'ce (2004).
\newblock Also in preprint \mbox{hep-ph/0411287}\relax
\relax
\bibitem{prep:97:31}
B.~Andersson \etal,
\newblock Phys.\ Rep.{} {\bf 97},~31~(1983)\relax
\relax
\bibitem{cpc:82:74}
T.~Sj\"ostrand,
\newblock Comp.\ Phys.\ Comm.{} {\bf 82},~74~(1994)\relax
\relax
\bibitem{cpc:69:155}
A.~Kwiatkowski, H.~Spiesberger and H.-J.~M\"ohring,
\newblock Comp.\ Phys.\ Comm.{} {\bf 69},~155~(1992)\relax
\relax
\bibitem{cpc:81:381}
K.~Charchula, G.A.~Schuler and H.~Spiesberger,
\newblock Comp.\ Phys.\ Comm.{} {\bf 81},~381~(1994)\relax
\relax
\bibitem{pr:d55:1280}
H.L.~Lai \etal,
\newblock Phys.\ Rev.{} {\bf D~55},~1280~(1997)\relax
\relax
\bibitem{tech:cern-dd-ee-84-1}
R.~Brun et al.,
\newblock {\em {\sc geant3}},
\newblock Technical Report CERN-DD/EE/84-1, CERN, 1987\relax
\relax
\bibitem{jhep:0602:032}
J. Pumplin \etal,
\newblock JHEP{} {\bf 0602},~032~(2006)\relax
\relax
\end{mcbibliography}
